\documentclass[acmsmall]{acmart} 

\def\BibTeX{{\rm B\kern-.05em{\sc i\kern-.025em b}\kern-.08emT\kern-.1667em\lower.7ex\hbox{E}\kern-.125emX}}
\makeatletter
\@namedef{r@tocindent4}{3pt}
\@namedef{r@tocindent5}{0pt}
\renewcommand\paragraph{\@startsection{paragraph}{4}{\z@}%
  {1.25ex \@plus1ex \@minus.2ex}%
  {-0em}%
  {\normalfont\normalsize\bfseries}}
\renewcommand\subsubsection{\@startsection {subsubsection}{3}{\z@ }%
  {-1.5ex\@plus -1ex \@minus -.2ex}%
  {1sp \@minus 0ex\nointerlineskip\vspace{-\lineskip}}%
  {\normalfont \normalsize \bfseries }}
\makeatother

\usepackage{multirow}
\usepackage{subfig}
\usepackage{adjustbox}
\usepackage{rotating}
\usepackage{lscape}
\usepackage{hyperref}
\usepackage{titlesec}

\usepackage{makecell}
\usepackage{comment}
\usepackage{caption}
\newcommand*{\q}[1]{\textcolor{red}{#1}}

\begin{document}

%
\title{Knowledge Tracing: A Survey }

%
\author{Ghodai Abdelrahman}
\email{ghodai.abdelrahman@anu.edu.au}
\author{Qing Wang}
\email{qing.wang@anu.edu.au}
\author{Bernardo Pereira Nunes}
\email{Bernardo.Nunes@anu.edu.au}
\affiliation{%
  \institution{School of Computing, 
 Australian National University}
  \city{Canberra}
  \state{ACT}
  \postcode{2601}
}

%

%
\begin{abstract}
Humans' ability to transfer knowledge through teaching is one of the essential aspects for human intelligence. A human teacher can track the knowledge of students to customize the teaching on students’ needs. With the rise of online education platforms, there is a similar need for machines to track the knowledge of students and tailor their learning experience. This is known as the \emph{Knowledge Tracing} (KT) problem in the literature. Effectively solving the KT problem would unlock the potential of computer-aided education applications such as intelligent tutoring systems, curriculum learning, and learning materials' recommendation. Moreover, from a more general viewpoint, a student may represent any kind of intelligent agents including both human and artificial agents. Thus, the potential of KT can be extended to any machine teaching application scenarios which seek for customizing the learning experience for a student agent (i.e., a machine learning model). In this paper, we provide a comprehensive and systematic review for the KT literature. We cover a broad range of methods starting from the early attempts to the recent state-of-the-art methods using deep learning, while highlighting the theoretical aspects of models and the characteristics of benchmark datasets. Besides these, we shed light on key modelling differences between closely related  methods and summarize them in an easy-to-understand format. Finally, we discuss current research gaps in the KT literature and possible future research and application directions.

\end{abstract}

%

%

%
\maketitle

\section{INTRODUCTION}
\label{Sec:IN}

Teaching is a vital activity to facilitate transfer of 
knowledge. It 
is well-known that one key factor of teaching is the ability of human teachers to track the learning progress of their students. This ability 
allows human teachers to adjust their teaching pace, materials and methodology 
to maximize the knowledge growth of each individual student. Over the past 30 years, a variety of online education platforms such as Massive Open Online Courses (MOOCs)~\cite{vardi2012will}, intelligent tutoring systems~\cite{psotka1988intelligent}, and educational games~\cite{long2017educational} have emerged to complement and sometimes completely replace conventional education systems. The COVID-19 pandemic, for example, has challenged conventional classroom-based teaching and sped up digital transformation in education systems. To alleviate the disruption of COVID-19, teachers and students around the world had to rapidly adjust to an online education mode. However, while there is a pressing need for teaching using computer technologies, technology-enhanced teaching has also posed new challenges, one of which is \emph{how to effectively track the learning progress of a student through their online interaction with teaching materials} -- a challenge known as the \emph{Knowledge Tracing} (KT) problem~\cite{anderson1986cognitive,BKT1992probabilistic,anderson1990cognitive}. In a nutshell, KT aims to observe, represent, and quantify a student's knowledge state, e.g., the mastery level of skills underlying the teaching materials. 

\begin{figure}[H]
\centering
\includegraphics[width=0.99\textwidth]{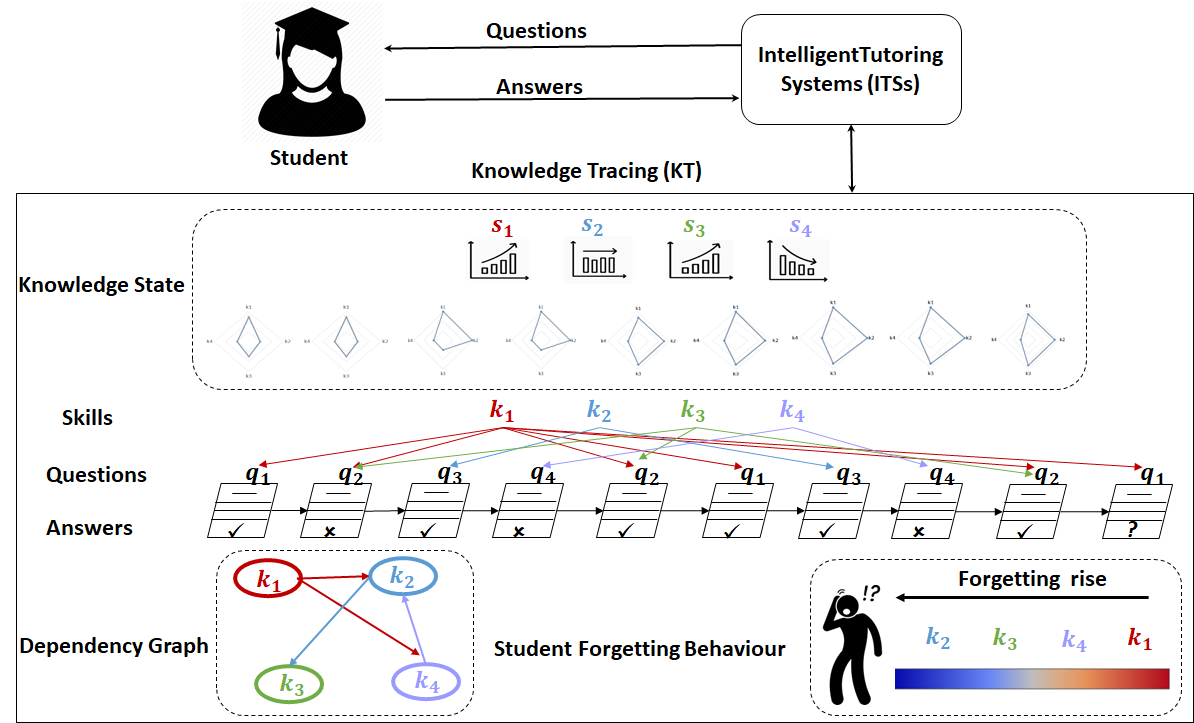}
\caption{An example scenario for knowledge tracing in an Intelligent Tutoring System (ITS).}
\label{fig:KT_Scenario}
\end{figure}


To better understand the KT problem, let us consider the learning activity depicted in Figure~\ref{fig:KT_Scenario}. It shows an interaction scenario between a student and an Intelligent Tutoring System (ITS) in which the student is given a sequence of questions taken from a question set $\{q_1,q_2,q_3,q_4\}$ and asked to answer these questions. During the interaction, the ITS 
estimates the student's knowledge states over the skills $\{k_1,k_2,k_3,k_4\}$ (e.g., math skills such as addition, subtraction and multiplication) that are required to answer these questions. However, capturing a student’s knowledge state is a challenging task due to several reasons: \smallskip
\begin{itemize}
\item Each question might require more than one \emph{skill}, which adds complexity to trace knowledge states. 
For instance, as shown by the arrows going from skills to questions in Figure~\ref{fig:KT_Scenario}, question $q_2$ requires the skills $\{k_1, k_3\}$. It is worth noting that \emph{skill} is also referred to as \emph{knowledge component} in some previous studies~\cite{liu2021survey}. \smallskip

\item Dependency among skills is another important factor to consider when tackling the KT problem. For example, although $q_3$ requires only skill $k_2$, skills $k_1$ and $k_4$ are prerequisites for skill $k_2$ according to the dependency graph shown in Figure~\ref{fig:KT_Scenario}. Thus, the mastery level of skills required by the question $q_3$ should also consider $k_1$ and $k_4$, in addition to skill $k_2$.\smallskip

\item A student's forgetting behavior~\cite{ebbinghaus2013memory} may result in decaying their knowledge over skills. By modeling forgetting features, skills can be ranked by their relevance to forgetting. For example, the bottom of Figure~\ref{fig:KT_Scenario} shows that skill $k_1$ is least affected by forgetting when the latest question $q_1$ is reached, whereas skill $k_2$ is the most affected one.\smallskip
\end{itemize}


Historically, the notion of knowledge tracing was introduced by Anderson et al. in a technical report~\cite{anderson1986cognitive} for cognitive modeling and intelligent tutoring in 1986, which was later published in the \emph{Artificial Intelligence} journal in 1990~\cite{anderson1990cognitive}. Since then, many attempts have been made to design machine learning models for solving the KT problem. Early attempts~\cite{BKT1992probabilistic, BKT2008more} followed Bayesian inference approaches, which usually relied on oversimplifying the model assumptions (e.g., assuming only one skill) to make the posterior computation tractable. Later, with the rise of classic machine learning methods such as logistic regression models~\cite{logreg95}, another direction followed by KT is to use parametric factor analysis approaches which trace a student's knowledge 
states and perform the answer prediction based on modeling a variety of factors~\cite{AFM06,AFM08, FactorsAnalysis2009}, including: (1) aspects about students such as prior knowledge, learning capacity, or learning rate; (2) aspects about learning materials such as familiarity, number of previous practices, or difficulty; (3) aspects about a learning environment itself such as the nature of the learning channel (paper- or computer-based) and the temporal context of the practice time (within an examination period or a regular study period). In addition to these, psychological studies about the learning behavior~\cite{learning_curve_theory} and forgetting behavior~\cite{Forgetting05} of students have also suggested additional factors, such as the time lapse between a student's different interactions and the number of times on practicing learning materials, to be considered when tracing knowledge states. It is worth noting that this direction of KT is still active~\cite{KTM19,KTM-DLF020} and is considered as an alternative to the recent state-of-the-art approaches based on deep learning.

Motivated by breakthroughs achieved by deep learning techniques~\cite{lecun2015deep}, 
deep learning KT models have emerged rapidly. Piech et al.~\cite{piech2015deep} has pioneered this direction of research and revealed the power of deep learning techniques for knowledge tracing. They proposed a model called {Deep Knowledge Tracing} (DKT) which applies Recurrent Neural Networks (RNNs)~\cite{lipton2015critical} to capture temporal dynamics in a sequence of interactions between questions and answers by a student, and based on that, to predict the student's answer for a new question. Empirical results showed that DKT outperformed traditional KT models on several benchmark datasets. This attempt highlighted the potential of using deep learning models in addressing the KT problem. In recent years, an increasing number of studies have exploited the development of deep learning KT models from different perspectives, 
including:
\begin{itemize}
    \item \textbf{Memory structures.~} Inspired by memory-augmented neural networks~\cite{graves2014neural,miller2016key}, deep learning KT models have been extended by augmenting more powerful memory structures, typically key-value memory, for capturing knowledge states dynamically at a finer granularity such as the mastery level of each individual skill (e.g.~\cite{zhang2017dynamic, SKVMN}).\smallskip
    \item \textbf{Attention mechanisms.~} Inspired by the \emph{Transformer} architecture~\cite{Attention_17} and some further developments in natural language processing applications, attention mechanisms have been incorporated into deep learning KT models for capturing the relationships among questions and their relevance to a student's knowledge states (e.g. ~\cite{SAKT19,AKT20,SAINT20,SAINT+21,pandey2020rkt}).\smallskip
    \item \textbf{Graph representation learning.~} Inspired by the representational power of graph learning techniques such as Graph Neural Networks \cite{GNN08,kipf_GCN_17}, deep learning KT models have been equipped with graph learning techniques to leverage the rich structural information from graphs that can flexibly model relationships among questions and skills (e.g.~\cite{GKT19,GIKT20,TongLHHCLM020}).\smallskip
    \item \textbf{Textual features.~} Question text may potentially contain a great wealth of information such as skills required by questions, difficulty of questions, and relationships between questions. Several deep learning KT models have leveraged textual features from question text for learning question representations and tracing a student's knowledge states (e.g.~\cite{YinLHCTW019,SuLLHYCDWH18,LiuHYCXSH21}).\smallskip

    \item \textbf{Forgetting features.~} Motivated by the \emph{learning curve theory}~\cite{learning_curve_theory}, a recent trend in developing deep learning KT models is to incorporate forgetting features so that a student's forgetting behavior can be taken into consideration for knowledge tracing (e.g.~\cite{wwwfOr19,for17,abdelrahman2021deep}).\smallskip

\end{itemize}
These studies facilitate a seamless translation of breakthroughs in deep learning techniques into the KT domain. 
So far, deep learning KT models have achieved the state-of-the-art results on the majority of benchmark datasets for knowledge tracing (a summary of the results obtained by different KT models is presented in Table \ref{tbl:ED}). 

\subsection{Contributions}

This paper performs a detailed survey that reviews, summarizes, classifies, and analyzes KT methods both from the traditional machine learning perspective and the recent deep learning perspective. It also presents KT benchmark datasets and applications. The main contributions are as follows:\smallskip

\begin{itemize}
    \item We highlight the key categories of KT methods and compare their architectures over multiple aspects including model design, knowledge state representation, assumptions for relationships between questions and skills, and consideration of student's forgetting behavior.\smallskip
    \item We present the chronic evolution for each KT category and discuss how each method is extended on the previous work.\smallskip
    \item We summarize the characteristics of well-known KT datasets and compare the performance of key KT methods on each dataset by consolidating results from the relevant literature.\smallskip
    \item We discuss application areas of KT that are not well explored currently to help derive future research directions in new venues.\smallskip
\end{itemize}

Note that a recent KT survey has been presented by Liu et al.~\cite{liu2021survey}. Despite their contributions to the field, our survey differs in depth and topics covered. The most notable differences are the comprehensive coverage of KT models; an important discussion on the differences in representation learning techniques related to key aspects of KT (such as knowledge state, forgetting behaviors, and knowledge components); an in-depth coverage of the KT datasets used by the relevant literature highlighting their similarities and issues; and, an extensive report of the results obtained by different KT models which allows future and accurate comparisons between them. 

\subsection{Scope and Structure}
This paper surveys 
the knowledge tracing literature to answer the following five research questions:\smallskip
\begin{itemize}
    \item[--] \textbf{[RQ1]}: What is the historical overview for the development of traditional KT techniques? \smallskip
    \item[--] \textbf{[RQ2]}: What are the key deep learning techniques that have been applied for solving the KT problem?\smallskip
    \item[--] \textbf{[RQ3]}: What are the datasets collected, pre-processed, and used for benchmarking KT tasks in the literature?\smallskip
    \item[--] \textbf{[RQ4]}: What are the different application areas that can benefit from the studies on KT?\smallskip
    \item[--] \textbf{[RQ5]}: What are the future research opportunities and challenges in KT?\smallskip
\end{itemize}

\begin{figure}[H]
\centering
\includegraphics[width=0.9\textwidth]{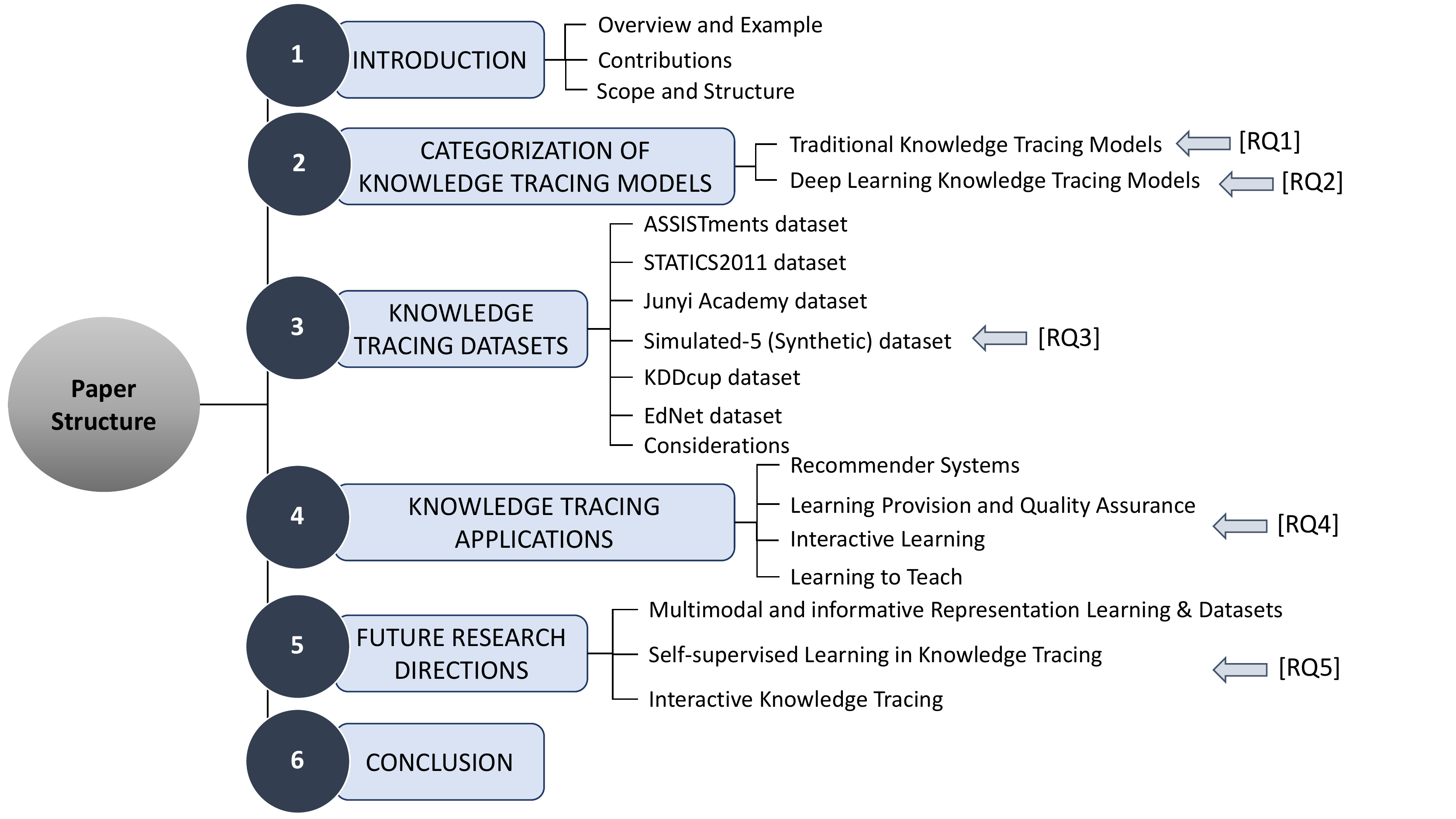}
\caption{Scope and structure of the survey.}
\label{fig:str}
\end{figure}

In the following sections, we discuss each of these five research questions in detail, as illustrated in Figure~\ref{fig:str}. To answer RQ1, we first survey the history of traditional KT techniques by summarizing the relevant studies and analysing their connections in Section~\ref{sec:RW}. Then, we investigate different types of deep learning techniques being proposed in the literature for solving the KT problem with respect to RQ2, which results in a taxonomy of deep learning KT models. For each type of deep learning techniques, we discuss their key assumptions and analyze their differences and connections. After that, in Section~\ref{Sec:DA}, we conduct a comprehensive analysis on the datasets in terms of data collection, pre-processing, characteristics, and the ground truth information, which answers RQ3. For RQ4, we discuss several types of KT applications in Section~\ref{sec:kt_apps}, particularly in relating to how KT techniques can enhance students' personalized learning experience and performance. Finally, we explore future research directions for KT which may enrich the field of study and provide a broad understanding of opportunities and limitations of existing KT models with respect to RQ5 in Section~\ref{sec:future} and conclude the paper.

\section{CATEGORIZATION OF KNOWLEDGE TRACING MODELS}
\label{sec:RW}

This section introduces a comprehensive categorization for the KT models according to the related works found in the literature. Generally speaking, there are two broad categories: (1) \emph{Traditional Knowledge Tracing Models}; and, (2) \emph{Deep Learning Knowledge Tracing Models}.

\subsection{Traditional Knowledge Tracing Models}
\label{sec:ch2_rl_pkt}
\begin{figure}[H]
\centering
\includegraphics[width=0.8\textwidth]{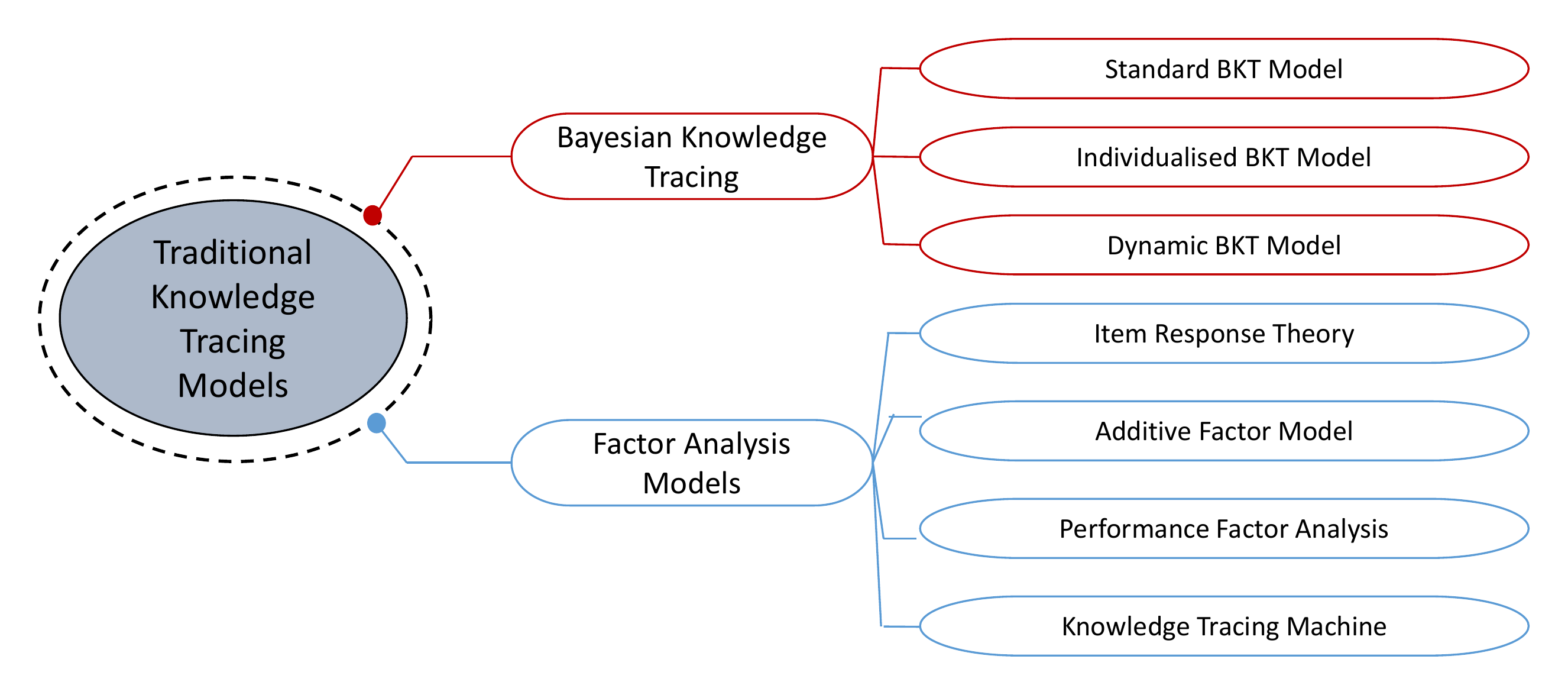}
\caption{An overview of traditional knowledge tracing models.}
\label{fig:BKT_part}
\end{figure}

Traditionally, there are two popular lines of research for knowledge tracing: the \emph{Bayesian Knowledge Tracing} and the \emph{Factor Analysis Models}. Figure~\ref{fig:BKT_part} provides an overview for major traditional knowledge tracing models that have been developed in the KT literature.

\subsubsection{\textbf{Bayesian Knowledge Tracing}} 

\emph{Bayesian Knowledge Tracing} (BKT) was motivated by the concepts of \emph{mastery learning}~\cite{mastery2000}.
Mastery learning assumes that all students can practise on a skill such that it may lead to mastery of that skill if two conditions are satisfied: (a) knowledge is appropriately described as a hierarchy of skills; and, (b) learning experiences are structured to ensure that students master skills lower than those higher in the hierarchy~\cite{corbett1994knowledge}.

\medskip
BKT models often use a probabilistic graphical model such as Hidden Markov Model~\cite{d2008more} and Bayesian Belief Network~\citep{villano1992probabilistic} to trace students' changing knowledge states as they practise skills. Central to these models is the Bayes' theorem that, for two events $A$ and $B$, the following holds:
\begin{equation}
    p(A|B)=\frac{p(B|A)p(A)}{p(B)}.
\end{equation}



In what follows we discuss the standard Bayesian approaches and their variations.

\begin{itemize}
    \item \textbf{Standard BKT Model}

The first BKT model was introduced by Corbett and Anderson in 1994 \citep{corbett1994knowledge}. The proposed model associates a skill with a binary knowledge state: $\{unlearned, learned\}$. This model only considers transitions from the unlearned state to the learned state, overlooking the forgetting theory (i.e., the probability of transition from a learned state to an unlearned state is always zero). Additionally, note that a student may make a mistake while in a learned state or guess correctly in an unlearned state.
We 
refer to this model as the \emph{standard BKT model} from now on.

\begin{figure}[H]
\begin{minipage}{0.8\textwidth}
\centering
\begin{tabular}{l| c }\hline
Parameter& Description\\\hline
 $p(L_0)$ & Probability of skill mastery by a student before learning \\ \hline
 $p(T)$ & Probability of transition from an unlearned state to a learned state  \\\hline  
 $p(S)$ & Probability of slipping by a student in a learned state  \\\hline
 $p(G)$ & Probability of guessing correctly by a student in an unlearned state \\\hline
\end{tabular}\captionof{table}{Four types of model parameters used in BKT}\label{tab:BKT-parameters}
\end{minipage}
\end{figure}
 There are two types of variables in the standard BKT model: (1) the \emph{Binary latent variables} which represent the knowledge states of a given student (i.e., a single variable per skill indicating \emph{learned state} and \emph{unlearned state}); and, (2) the \emph{Binary observed variables} which represent how students attempt questions (i.e., a variable per question indicating whether a question is answered correctly or not). 

 Figure~\ref{fig:BKT} shows four types of model parameters in the standard BKT model. For each skill, there is one set of four corresponding parameters. At each time step $n\geq 1$, the model estimates the probability $p(L_n)$ of skill mastery by a student by:
 
\begin{equation}
    p(L_{n})=Posterior(L_{n-1})+(1-Posterior(L_{n-1}))\ast p(T),
\end{equation}

\noindent where Posterior$(L_{n-1})$ is the posterior probability of being in a learned state given the observation to the $n$-th attempt by a student, calculated as:

\[\hspace{-8.5cm}Posterior(L_{n-1})=\]
\[\hspace{2cm}
\begin{cases}
    \dfrac{p(L_{n-1})\ast(1-p(S))}{p(L_{n-1})\ast(1-p(S))+(1-p(L_{n-1}))\ast p(G)} \text{ if the $n$-th attempt is \emph{correct};}\\
    \dfrac{p(L_{n-1})\ast p(S)}{p(L_{n-1})\ast p(S)+(1-p(L_{n-1}))\ast(1-p(G))} \text{ otherwise.}
\end{cases}
\]
\medskip

Hence, the probability of a student to correctly answer a question at each time step $n$ is the sum of the probability of either mastering the skill but making a ``slip'' or not mastering the skill but making a correct ``guess'', as formulated by:
\begin{equation}
    p(L_{n})\ast(1-p(S))+(1-p(L_{n}))\ast p(G)).
\end{equation}
\smallskip

 Figure~\ref{fig:BKT}a shows the standard BKT model with one skill node $k$. Starting with the prior probability $p(L_{0})$ of skill mastery, the latent variable for skill $k$ is transitioned from one time step $t-1$ to the next time step $t$ based on the probability $p(T)$. The corresponding observed variable $y$ represents the answer node, i.e., how a student attempts questions that require skill $k$, which is based on the probabilities $p(G)$ and $p(S)$. 
 
 \medskip

\begin{figure}[t!]
\centering
\includegraphics[width=0.7\textwidth]{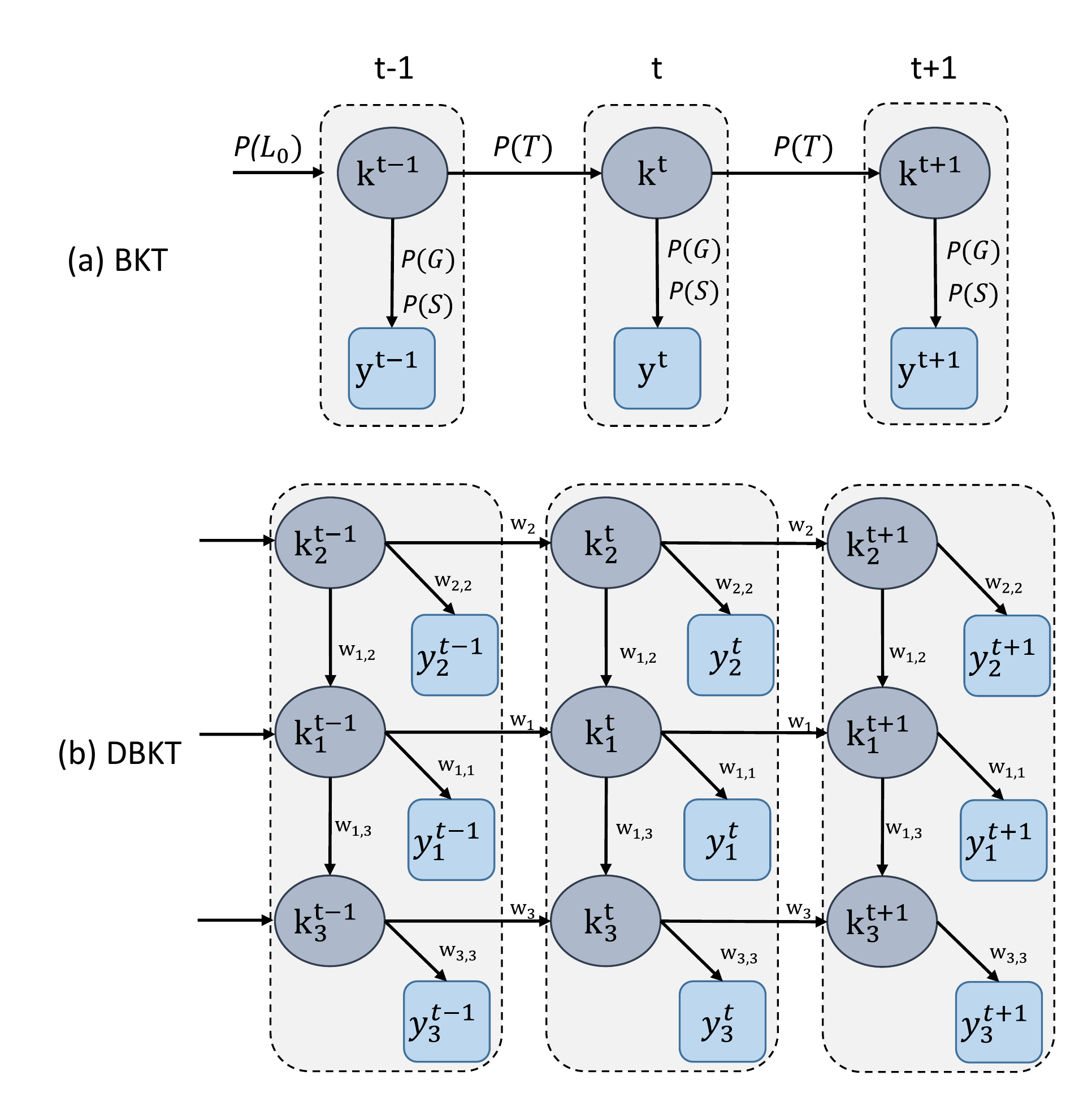}
\caption{Comparing the model architectures between (a) BKT and (b) DBKT, where latent variables for skills at the time step $t$ are denoted as $k_i^t$ and their correspondingly observed variables for answers at the time step $t$ are denoted as $y^t_i$.}
\label{fig:BKT}
\end{figure}

 
\item \textbf{Individualized BKT Model}

  One limitation of the standard BKT model is that it has no model parameters specific to students. All students are assumed to have the same prior knowledge and the same learning rate for any skill. As a result, the standard BKT model may underestimate the learning performance for above-average students, but overestimate the learning performance for below-average students \citep{corbett1994knowledge}. 
  
  \medskip
  
 
  
To alleviate this limitation, several attempts  \citep{corbett1994knowledge,Pardos_2010,lee2012impact,Yudelson_2013, khajah2014} have been made to extend the standard BKT model by introducing student-specific parameters. Corbett and Anderson \citep{corbett1994knowledge} considered to add  individual weights for each student, one weight corresponding to each of the four parameter types in the standard BKT model. Pardos and Heffernan~\citep{Pardos_2010} focused on individualizing the prior probability of skill mastery $p(L_0)$ heuristically for each student. Lee and Brunskill~\cite{lee2012impact} also individualized the four parameter types in the standard BKT model for students; nonetheless, the combined effects of skill-specific and student-specific parameters were unexplored. Yudelson et al.~\citep{Yudelson_2013} introduced an approach for individualizing BKT models that account for student differences with respect to two types of model parameters, the prior probability of skill mastery $p(L_0)$ and the probability of transition $p(T)$. The idea was to first define student-specific parameters and skill-specific parameters. Then, the gradients were explicitly computed in terms of both student-specific and skill-specific parameters. The underlying Hidden Markov Model remains unchanged. It turns out that adding student-specific parameter for $p(T)$ is more beneficial for the model accuracy than adding student-specific parameter for $p(L_0)$. Later, Khajah et al.~\cite{khajah2014} proposed to extend the standard BKT model by personalizing the guess and slip probabilities $p(G)$ and $p(S)$ based on student ability and problem difficulty.

\medskip

Compared to the standard BKT model, the individualized models can provide better correlation between actual and expected accuracy among students, leading to more effective decisions or improving the accuracy of predicting student performance. Table \ref{tab:IBKT-parameters} summarizes the skill- and student-specific parameters used in the individualized BKT models.

\begin{table}[t!]
\centering
\begin{tabular}{l| cc | cc | cc | cc | cc }\hline
\multirow{2}{*}{Parameter}  & \multicolumn{2}{c|}{\citep{corbett1994knowledge}} &\multicolumn{2}{c|}{\citep{Pardos_2010}} &\multicolumn{2}{c|}{\cite{lee2012impact}} &\multicolumn{2}{c|}{\citep{Yudelson_2013}} & \multicolumn{2}{c}{\citep{khajah2014}}\\\cline{2-11}
   & skill & student &skill & student &skill & student &skill & student  &skill & student\\\hline
 $p(L_0)$ & $\checkmark$&$\checkmark$ & $\checkmark$&$\checkmark$ &-&$\checkmark$& $\checkmark$&$\checkmark$&$\checkmark$ &- \\\hline
 $p(T)$ &$\checkmark$ & $\checkmark$ &$\checkmark$ &-&-&$\checkmark$& $\checkmark$&$\checkmark$&$\checkmark$ &-\\\hline  
 $p(S)$ & $\checkmark$& $\checkmark$ &$\checkmark$ &-&-&$\checkmark$& $\checkmark$& -&$\checkmark$& $\checkmark$\\\hline
 $p(G)$ &$\checkmark$ & $\checkmark$ &$\checkmark$ &-&-&$\checkmark$& $\checkmark$& -&$\checkmark$& $\checkmark$\\\hline
\end{tabular}\captionof{table}{Model parameters of individualized BKT models}\label{tab:IBKT-parameters}
\end{table}



 
\medskip
\item \textbf{Dynamic BKT Model}

In the early years, the BKT models have assumed that each question requires only one skill and different skills are independent from each other \citep{corbett1994knowledge,Pardos_2010,lee2012impact,Yudelson_2013}. Thus, these models cannot handle questions that require multiple skills, nor represent relationships between different skills. To address this limitation, K\"{a}ser et al.~\cite{DBN} proposed to jointly model multiple skills and dependencies between different skills using \emph{Dynamic Bayesian Network} (DBN). They aimed to capture prerequisite skill hierarchies within a single model, e.g., one skill is conditionally dependent on another skill if the former is a prerequisite for mastering the latter. Similar to the standard BKT model, DBN considers the same two types of variables: binary latent variables and binary observed variables.


\medskip
At each time step, a latent variable for each skill is associated with an observed variable. A forget probability $p(F)$ is introduced, in addition to the four types of model parameters $\{p(L_0), p(T), p(S), p(G)\}$ in the standard BKT model. Dependencies between different skills are learnt as weights $\textbf{w}$ of a log-linear model. Let $f : X\times O \rightarrow \mathbb{R}^d$ denote a mapping from a latent space $X$ and an observed space $O$ to d-dimensional feature vectors, and $c$ be a normalizing constant. The objective of the log-linear model is to find the model parameters $\{p(L_0), p(T), p(S), p(G), p(F), \textbf{w}\}$ that maximize the likelihood of the joint probability of $x_i\in X$ and $y_i\in O$ as formulated below:
\[
L(\textbf{w})=\sum_i ln \big(\sum_{x_i} exp(\textbf{w}^\intercal f(x_i, y_i)-c) \big).
\]

 Figure~\ref{fig:BKT}b shows the Dynamic Bayesian Network (DBKT) with three skill nodes (denoted as $k_1$, $k_2$ and $k_3$). As indicated by the directed arrows, the latent variable for skill $k_2$ depends on the latent variable for skill $k_1$, and the latent variable for skill $k_1$ depends on the latent variable for skill $k_3$. Further, at each time step $t$, latent variables for skills depend on their latent variables in the previous time step $t-1$, while $y_i$ is the corresponding observed answer nodes.  

\end{itemize}

\subsubsection{\textbf{Factor Analysis Models}}

Factor analysis models are theoretically supported by the \emph{Item Response Theory} (IRT)~\cite{embretson2013item}, which has played a large role in educational assessment and measurement.
The key idea is to estimate student performance by learning a function, usually a logistic function, based on various factors in a population of students who solve a set of problems. It is important to note that, although an item in the original IRT corresponds to a question involving a single skill, later works in this line have been generalized to considering an item that may involve multiple skills. The mapping between items and skills is often represented in the form of Q-matrix, i.e., an entry $q_{jk}$ in a Q-matrix is 1 if an item $j$ involves a skill $k$; or 0, otherwise. Q-matrix is commonly assumed as the side information. \smallskip

\begin{itemize}
    \item \textbf{Item Response Theory (IRT)}
    
The history of IRT can be traced back to Thurston's pioneering work in the 1920s \cite{lord1968statistical} and several other works in the 1950s and 1960s~\cite{IRT1951,IRT1969,David,Lord}. IRT is built upon the following assumptions: (a) The probability that a student correctly answers an item can be formulated as an \emph{item response function} based on the parameters of the student and the item; (b) The item response function monotonically increases with respect to the ability $\theta_{i}$ of a student $i$; (c) For a student with the ability $\theta_{i}$, items are considered conditionally independent.

 The \emph{Rasch} model~\cite{rasch1993probabilistic} is often refereed to as the simplest IRT model, in which the item response function is defined by a one-parameter logistic regression (1PL) model. Let $\mathcal{L}(\cdot)$ be a logistic function. By taking into account a difficulty parameter $b_j$ that models the difficulty of an item $j$, the probability $p_{ij}$ that a student $i$ correctly answers an item $j$ is defined as:
\begin{equation}
 p_{ij}=\mathcal{L}\Big(\theta_{i}-b_{j}\Big)=\frac{exp^{(\theta_{i}-b_{j})}}{1+exp^{(\theta_{i}-b_{j})}}.
\end{equation}

Several multiple parameter logistic regression models have also been developed for IRT, e.g., a four-parameter logistic (4PL) model introduced by Barton and Lord \cite{barton1981upper}:
\begin{equation}
    p_{ij}=c_{j}+(d_{j}-c_{j})\mathcal{L}\Big(a_j(\theta_{i}-b_{j})\Big)=c_{j}+(d_{j}-c_{j})\frac{exp^{a_{j}(\theta_{i}-b_{j})}}{1+exp^{a_{j}(\theta_{i}-b_{j})}},
\end{equation}
\smallskip

\noindent where $a_{j}$ is a discrimination parameter to model how well an item $j$ can differentiate students, $c_{j}$ is a guessing parameter to model the effect of guessing, and $d_{j}$ is a slipping parameter to model the effect of careless errors.
 
 \medskip

IRT has been extended in many different directions. Wilson et al.~\cite{IRT2016back} proposed \emph{Hierarchical
IRT} (HIRT) and \emph{Temporal IRT} (TIRT). HIRT exploits structure among questions by assuming that related questions (i.e., questions sharing similar skills) have difficulty parameters drawn from the same distribution. Different questions might vary in difficulty, but questions for trivial skills tend to be easier while ones for difficult skills tend to be harder. TIRT models each parameter in the logistic model (e.g., 4PL) as a time-varying stochastic process such as a \emph{Wiener random} process~\cite{WP2010}.

\medskip

\item \textbf{Additive Factor Model (AFM)}

The \emph{Additive Factor Model}~\cite{AFM08}, originated from \emph{Learning Factors Analysis} (LFA)~\cite{AFM06}, is a logistic regression model under four assumptions: (1) the prior knowledge of students may vary; (2) students learn at the same rate; (3) some skills are more likely to be known than others; and, (4) some skills are easier to be learnt than others. In this model, a difficulty parameter $\beta_k$ and a learning rate parameter $\gamma_k$ are assigned for each skill $k$, respectively. 
\medskip
 
The key idea of AFM is that the probability of answering an item correctly by a student is proportional to an additive combination of the ability of the student, the difficulty of skills involved in the item, and the amount of learning gained from each attempt. Let $K(j)$ be the set of skills involved in an item $j$, which can be obtained from a Q-matrix, and $T_{ik}$ be the number of times that a student $i$ has attempted on an item involving a skill $k$. AFM defines the probability of answering an item correctly by a student $i$ on an item $j$ as:

\begin{equation}
    \label{AFM}
    p_{ij}=\mathcal{L}\Big(\theta_{i}+\sum_{k\in K(j)}(\beta_{k}+\gamma_{k}T_{ik})\Big).
\end{equation}

\medskip

\item \textbf{Performance Factor Analysis (PFA)}

 \emph{Performance Factor Analysis} (PFA)~\citep{FactorsAnalysis2009} overcomes the limitation of AFM which ignores the evidence of learning in the successful and unsuccessful attempts on items by a student. 
 \medskip

 The key idea of PFA is to discard the parameter $\theta_{i}$ used in the previous models and instead count for success and unsuccessful attempts separately. PFA has three parameters for each skill, including: (1) $\beta_k$ is for the difficulty of a skill $k$; (2) $\gamma^S_{k}$ is for the effect of learning a skill $k$ after successful attempts; and, (3) $\gamma^F_{k}$ is for the effect of learning a skill $k$ after unsuccessful attempts. Conceptually, $\gamma^S_{k}$ and $\gamma^F_{k}$ reflect the learning rate for a skill $k$ when being applied successfully and unsuccessfully.
 
 \medskip
Let $T^S_{ik}$ and $T^F_{ik}$ be the number of successful attempts and the number of unsuccessful attempts made by a student $i$ on a skill $k$, respectively. Then PFA calculates the probability that a student $i$ correctly answer an item $j$ as:

\begin{equation}
    p_{ij}=\mathcal{L}\Big(\sum_{k\in K(j)}(\beta_{k}+\gamma^S_{k}T^S_{ik}+\gamma^T_{k}T^F_{ik})\Big).
\end{equation}
\smallskip

\item \textbf{Knowledge Tracing Machine (KTM)}

\emph{Knowledge Tracing Machine} (KTM) was recently proposed by Vie and Hisashi~\cite{KTM19}, which generalizes \emph{Factorization Machine} (FM)~\cite{FM12,FM16} for student modeling. 
\medskip

KTM allows to consider an arbitrary number of factors about students, items, skills, successful and unsuccessful attempts, or extra information about the learning environment, such as using a mobile or a laptop. For a number $N$ of factors, we denote all factors involved in an event by a sparse vector $x$ of length $N$ such that $x_{k} > 0$ if a factor $k\in [1, N]$ is involved in the event; or 0, otherwise. Then KTM estimates the probability $p_{ij}$ of a correct answer on an item $j$ by a student $i$ with an event involving $x$ as follows:
\begin{equation}
    \label{KTM}
    p_{ij}=\mathcal{L}\Big(\sum_{k=1}^{N} w_{k}x_{k}+\sum_{1 \leqslant k< l \leqslant N} x_{k}x_{l}\langle v_{k},v_{l}  \rangle +\mu\Big)
\end{equation}

\noindent where $\mu$ is a global bias, and each factor $k$ is modeled by both a weight $w_{k} \in \mathbb{R}$ and an embedding vector $v_{k} \in \mathbb{R}^{d}$ for some dimension $d$. The first term models the logistic regression of all factors and the second term models pairwise interactions between different factors. When $\mathcal{L}$ is a logistic function, KTM include IRT, AFM and PFA as special cases~\cite{KTM19}.

\medskip
Wenbin et al.~\cite{KTM-DLF020} proposed \emph{Knowledge Tracing Machine by modeling cognitive item Difficulty and Learning and Forgetting (KTM-DLF)} that extends KTM by adding factors related to the forgetting behavior of students. They represented the forgetting by time lapse since the last successful attempt for the involved skills.

\end{itemize}

\subsubsection{Discussion}
Both Bayesian knowledge tracing and factor analysis models have strengths and weaknesses. We discuss their connections and differences from three aspects: model parameters, model inference, and temporal analysis.

\begin{itemize}
    \item \textbf{Model parameters:} Most recent models in BKT and FAM have taken into account both student- and skill-specific model parameters. Early BKT models were primarily centered on the four parameters: prior learning parameter $p(L_0)$, learning rate parameter $p(T)$, guess parameter $p(G)$, and slip parameter $p(S)$ \citep{corbett1994knowledge} and their student-specific variants. These early BKT models usually assume that there is no forgetting parameter. Only some recent works have incorporated the forgetting parameter~\cite{DBN}. For factor analysis models, most of them have been developed with similar or more flexible model parameters than BKT models. Particularly, recent works such as KTM~\cite{KTM19} consider a wide range of factors, enabling a flexible way to incorporate the side information into student modelling. 
    
    \medskip
    \item \textbf{Model inference:} To keep the Bayesian inference tractable, BKT models typically assume a first-order Markov chain when making inference based on a sequence of past question-answering history, i.e., only considering the most recent observation. This assumption however limits their ability to model complex dynamics on student learning behaviors. Some recent dynamic BKT models such as DBN~\cite{DBN} are often computationally intractable, and thus they trade-off predication accuracy for computational efficiency. On the other hand, factor analysis models usually do not explicitly make inferences on knowledge states of a student (e.g., decide whether a student has achieved a certain level of skill mastery 
    by tracing knowledge states). Instead, they target to maximize other model parameters such as the learning rate.
 
 \medskip
    \item \textbf{Temporal analysis:} BKT models essentially deal with a sequence prediction problem based on the history of student learning. In contrast, factor analysis models do not consider the order of questions in which a student's answers are observed. For example, given two questions and their corresponding answers from a student, whether one question is answered before the other is not important for factor analysis models. Nonetheless, by incorporating extra temporal features of student learning behaviors, factor analysis models can be enhanced to analyze temporal aspects of student learning. 
\end{itemize}\smallskip

A number of attempts have been made to leverage the best from both worlds of Bayesian knowledge tracing and factor analysis models. For example, ~\cite{IRT2014general,IRT2014BKT,IRT2014BKT2} extended the IRT using the Bayesian inference to customize the estimation of question difficulty based on observations of each student. Further work has been done through incorporating factors that reflect the characteristics of individual students~\citep{d2008more,pardos2010modeling}, or the characteristics of specific items of assessment within skills~\citep{pardos2011kt}.



\subsection{Deep Learning Knowledge Tracing Models}
\label{sec:ch2_rl_dkt}
\begin{figure}[t!]
\centering
\includegraphics[width=0.99\textwidth]{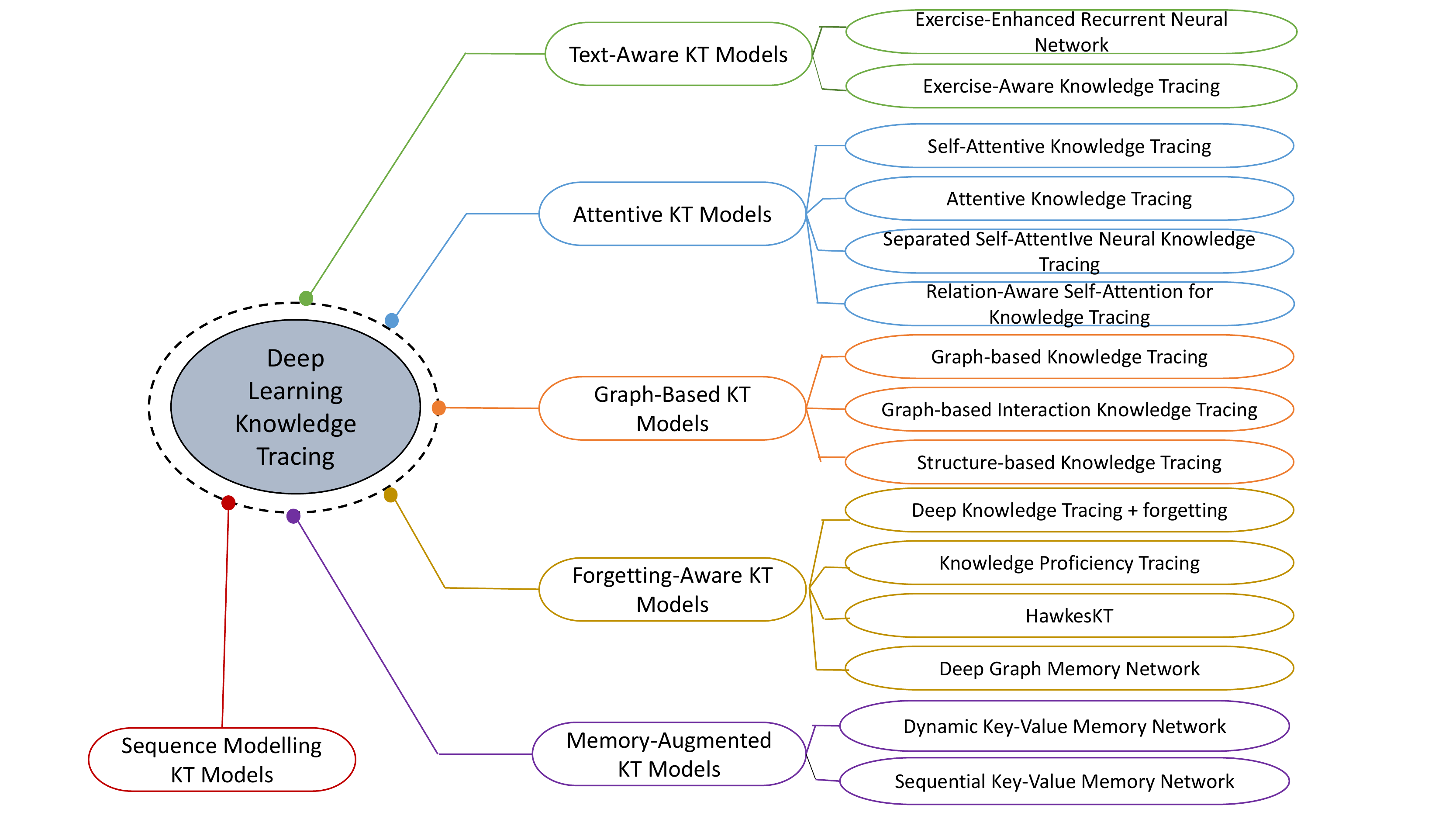}
\caption{A taxonomy of deep learning knowledge tracing models.}
\label{fig:KT_DL}
\end{figure}

Inspired by the success of deep learning~\citep{SCHMIDHUBER201585,NIPS2014_5346,Graves2013,2013arXiv1308.0850G,lecun2015deep}, recent researches on knowledge tracing have applied deep learning techniques. 
Figure~\ref{fig:KT_DL} presents a taxonomy of deep learning knowledge tracing models. 

\subsubsection{Sequence Modeling for Knowledge Tracing}

A knowledge tracing task is typically modeled as a sequence prediction problem from a machine learning perspective. Let ${Q}=\left\{ q_{1},\ldots,q_{|Q|}\right\}$ be the set of all distinct questions in a dataset. Each  ${q_{i}\in Q}$ may have a different level of difficulty, which is not explicitly provided. When a student interacts with the questions in ${Q}$, a sequence of interactions ${X}=\langle  x_{1},x_{2},\ldots,x_{t-1}\rangle $ undertaken by the student can be observed, where $x_i=\left(q_{i},y_{i}\right)$ consisting of a question ${q_{i}}$ and answer $y_{i}\in\{0,1\}$. $y_i=0$ means that $q_{i}$ is incorrectly answered and $y_i=1$ means that $q_{i}$ is correctly answered. 

\begin{definition}
Given a sequence of interactions $X$ that contains the previous question answering of a student, the \emph{knowledge tracing} problem is to predict the probability $p_t$ of correctly answering a new question $q_t$ at the time step $t$ by the student, i.e., ${p_{t}=\left(y_{t}=1|q_{t},{X}\right)}$.
\end{definition}

\emph{Deep Knowledge Tracing} (DKT)~\cite{piech2015deep} pioneered the use of deep learning for knowledge tracing. It employs \emph{Recurrent Neural Network} (RNN)~\citep{lipton2015critical} and a Long Short Term Memory (LSTM)~\citep{LSTM} to predicate the probability of correctly answering a question at each time step. 
A sequence of hidden states $\langle h_{1}, h_{2},\ldots, h_{n} \rangle$ is computed which encodes the sequence information obtained from previous interactions.
At each time step $t$, the model calculates the hidden state $h_t$ and the student's response $p_t$ as follows:
\begin{equation}
    h_t=Tanh(W_{hx}x_{t}+W_{hh}h_{t-1}+b_{h})
\end{equation}
\begin{equation}
    p_{t}=\sigma(W_{hy}h_{t}+b_{p})
\end{equation}

\smallskip
\noindent where the ${Tanh(u_{i})=(e^{u_{i}}-e^{-u_{i}})/(e^{u_{i}}+e^{-u_{i}})}$ and ${\sigma (u_{i})=1/(1+e^{-u_{i}})}$ are activation functions, $W_{hx}$, $W_{hh}$ and $W_{hy}$ are weight matrices, and $b_h$ and $b_p$ are bias vectors.

Despite the promising performance, DKT has several limitations. First, it assumes only one hidden KC (i.e., a skill) in a student's knowledge state $h_{t}$. Second, it cannot model the relationships among multiple KCs. Third, it assumes that all questions are equally likely related to each other, which may not hold in many scenarios as some questions may be more relevant to each other than the remaining of the sequence. Thus, various attempts have been made on extending DKT with the aim of enhancing the model capacity for tackling the KT problem. Below, we review the related work in the following areas of extension.

A number of KT models have extended DKT~\cite{piech2015deep} to address its limitations. For example, Xiong et al.~\cite{Going16} proposed \emph{Extended-Deep Knowledge Tracing} that extends DKT by adding auxiliary student features such as previous knowledge, question answering rates and time spent on learning and practice; and, exercise features, such as textual information, question difficulty, skill hierarchies and skill dependencies. A variant of DKT, called (DKT+)~\cite{YeungY18}, was proposed to augment the original DKT loss function with two additional regularization terms to address the limitations in DKT's ability to reconstruct an answer input and reduce inconsistency of answer prediction for questions sharing similar KCs. Minn et al.~\cite{DKT-DSC18} proposed an extension of DKT, named \emph{Deep Knowledge Tracing with Dynamic Student Classification} (DKT-DSC), that uses K-means to cluster student profiles into groups based on their performance over the KCs and dynamically update the current cluster information over time while the performance changes.

\subsubsection{Memory-Augmented Knowledge Tracing Models}

To trace complex KCs learned by students, several works have extended DKT by augmenting an external memory structure, inspired by memory-augmented neural networks~\cite{graves2014neural}. In particular, following \emph{Key-Value Memory Network} (KVMN)~\citep{miller2016key}, a key-value memory has been employed to represent knowledge state, which has more representational power than a hidden variable used in DKT. Such a key-value memory consists of two matrices: \emph{key} and \emph{value}. The \emph{key} matrix stores the representations of KCs and the \emph{value} matrix stores the student’s mastery level of each KC. Below, we discuss two popular key-value memory networks for knowledge tracing.

\begin{figure}[H]
\centering
\vspace*{0cm}
\includegraphics[width=0.85\textwidth]{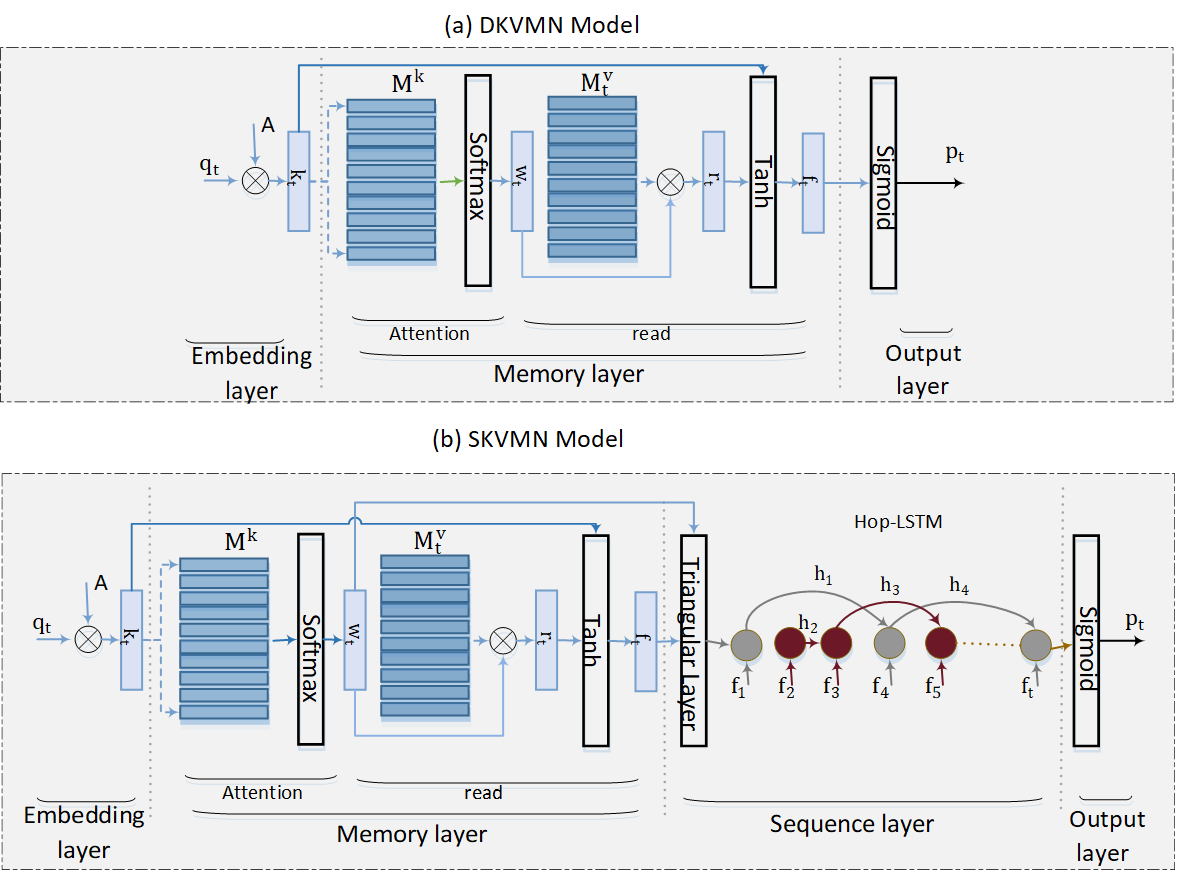}
\caption{A comparison of the model architecture design between (a) DKVMN and (b) SKVMN.}
\label{fig:kvm}
\end{figure}

\begin{itemize}
    \item \textbf{Dynamic Key-Value Memory Network (DKVMN)~}
    
\emph{Dynamic Key-Value Memory Network} (DKVMN)~\cite{zhang2017dynamic} has augmented DKT with two memory matrices: \emph{key} and \emph{value}. To trace how the knowledge state of a student evolves over time, unlike KVMN in which both \emph{key} and \emph{value} matrices are static~\citep{miller2016key}, DKVMN designs the \emph{value} matrix to be dynamic while remaining the \emph{key} matrix to be static.\smallskip

Figure~\ref{fig:kvm}a shows the model architecture of DKVMN, where $\mathbf{M}^{k}\in\mathbb{R}^{N\times d_k}$ is the \emph{key} matrix and $\mathbf{M}_t^{v}\in\mathbb{R}^{N\times d_v}$ is the \emph{value} matrix at the time step $t$. It is assumed that there are $N$ latent KCs underlying all questions in a learning task. For a question $q_{t}$ at the time step $t$, a correlation weight 
$w_{t}$ is computed, which represents the correlation between the question $q_{t}$ and the underlying latent KCs stored in the \emph{key} matrix $\mathbf{M}^{k}$. The model first retrieves a student's knowledge state with regard to the question $q_{t}$ from the \emph{value} matrix $\mathbf{M}_{t}^{v}$, calculated as: 
\begin{equation}
\label{eq:readVec}
r_{t}=\sum_{i=1}^{N}w_{t}(i)\mathbf{M}_{t}^{v}(i).
\end{equation}
Then, the student's response for the question $q_t$ is predicted based on the retrieved knowledge state. After the student answers the question $q_t$, the \emph{value} matrix is updated to reflect the knowledge growth of the student after working on $q_t$.

\medskip

\item\textbf{Sequential Key-Value Memory Network (SKVMN)~}

\emph{Sequential Key-Value Memory Network} (SKVMN)~\cite{SKVMN} aimed to address a limitation in DKT and DKVMN that KCs required by answering the past questions in a sequence are not necessarily relevant to KCs required by answering the current question. Thus, SKVMN employs a modified LSTM for sequential modeling, called \emph{Hop-LSTM}, while remaining the same key-value memory structure and loss function as in DKVMN. Different from the standard LSTM, Hop-LSTM can explicitly capture sequential dependencies among questions in a sequence of interactions, and update the knowledge state of a student based on their responses to relevant questions. \smallskip

Figure~\ref{fig:kvm}b shows the model architecture of SKVMN. More precisely, two LSTM cells in Hop-LSTM are connected only if the input question of one LSTM cell is sequentially dependent on the input question of the other LSTM cell. This means that Hop-LSTM has the capability of hopping across the LSTM cells when their input questions are irrelevant to the current question ${q_{t}}$. Thus, it enables to capture long-term dependencies among questions that require similar KCs. 
\end{itemize}

\subsubsection{Attentive Knowledge Tracing Models}


Following the \emph{Transformer} architecture~\cite{Attention_17}, several works~\cite{SAKT19,AKT20,SAINT20,SAINT+21,pandey2020rkt} have attempted to incorporate an attention mechanism into KT models. Although the attention mechanisms introduced by these works vary, their key ideas are similar, i.e., to learn the attention weights of questions in a sequence of interactions in a way that can reflect the relative importance of these questions for predicting the probability of correctly answering the next question. This mitigates one limitation of DKT that treats all questions in a sequence of interactions equally important. In what follows, we discuss the main attentive knowledge tracing models. Table~\ref{tab:AMKT} briefly summarizes these models. 

\begin{table}[ht]
    \centering
    \begin{tabular}{ l| c c |l}\hline
 \textbf{KT Model} & \textbf{Forgetting-Aware} & \textbf{Text-Aware} &  \hspace{0.3cm}\textbf{Attention Mechanism}\\ \hline
         SAKT~\cite{SAKT19}  & $\times$& $\times$ &  Multi-head self-attention\\
 AKT~\cite{AKT20} & $\checkmark$ & $\times$ &  Monotonic attention\\  
 SAINT~\cite{SAINT20}& $\times$ & $\times$ & Multi-head self-attention\\
 SAINT+~\cite{SAINT+21}&$\checkmark$& $\times$ &Multi-head self-attention\\
  RKT~\cite{pandey2020rkt} & $\checkmark$ & $\checkmark$ & Relational multi-head self-attention\\\hline
    \end{tabular}
    \caption{A comparison of attentive knowledge tracing models.}
    \label{tab:AMKT}
\end{table}

\begin{itemize}
\item \textbf{Self-Attentive Knowledge Tracing (SAKT)}

\emph{Self-Attentive Knowledge Tracing} (SAKT)~\cite{SAKT19} was the first to add an attention mechanism into the KT models. It uses the scaled dot-product attention mechanism proposed by Vaswani et al.~\cite{Attention_17} to learn attention matrices using multiple attention heads. Specifically, each attention matrix contains relative weights from a representative subspace, which indicate the importance of questions in the past interactions
for predicting a student's answer to the current question. Then, attention matrices from
different representative subspaces are sent to a feed forward network for predicating student performance.

\medskip
\item \textbf{Attentive Knowledge Tracing (AKT)}

\emph{Attentive Knowledge Tracing} (AKT) was proposed by Ghosh, Heffernan, and Lan~\cite{AKT20}. AKT differs from SAKT in its attention mechanism called \emph{monotonic attention} (i.e., a modified, monotonic version of the scaled dot-product attention mechanism~\cite{Attention_17}) that can reduce attention weights for questions in a sequence of interactions proportional to their time distance in an exponential decay rate. The exponential weight decay is meant to consider the forgetting effect in a student's memory over time. In addition, an embedding representation was proposed to take into account a parameter for controlling how far a question deviates from a knowledge component it involves by following the \emph{Rasch} model~\cite{rasch1993probabilistic}. 
\medskip

\item\textbf{Separated Self-AttentIve Neural
Knowledge Tracing (SAINT)}

\emph{Separated Self-AttentIve Neural
Knowledge Tracing} (SAINT), proposed by Choi et al.~\cite{SAINT20}, differs from AKT and SAKT in the way that it has further applied a encoder-decoder model along with the scaled dot-product
attention mechanism as in the original architecture of Transformer~\cite{Attention_17}. Specifically, SAINT separates a sequence of interactions by a student into a \emph{question embedding sequence} and a \emph{response embedding sequence}, which are then sent to the encoder and the decoder as input, respectively. The encoder and decoder are combinations of multi-head attention networks with the scaled dot-product
attention mechanism~\cite{Attention_17}.
\smallskip

Recently, SAINT was extended by adding two time-related features into a response embedding sequence:
\emph{elapsed time} for the time taken by a student to answer each question, and \emph{lag time} for the time interval between two consecutive learning interactions. This variant was named as the SAINT+ model~\cite{SAINT+21}. 
\medskip

\item \textbf{Relation-Aware Self-Attention for Knowledge Tracing (RKT)}

\emph{Relation-Aware Self-Attention for Knowledge Tracing} (RKT) was proposed by Pandey and Srivastava~\cite{pandey2020rkt}. Similar to SAKT and SAINT, RKT employs the scaled dot-product
attention mechanism proposed by Vaswani et al.~\cite{Attention_17} to learn attention weights using multiple attention heads. However, differs from the other attention-based KT models, RKT combines attention weights with \emph{relation coefficients}, which are obtained from \emph{exercise relation modeling} and \emph{forgetting behavior modeling}. For the exercise relation modeling, it leverages the text information of questions (e.g., a question's textual information) to represent questions and estimate the relation between questions in a sequence of past interactions. For the forgetting behavior modeling, similar to AKT, RKT considers an exponential decay to count for a student's forgetting behavior over time.
\end{itemize}

\subsubsection{Graph-Based Knowledge Tracing Models}
In KT tasks, various relational structures often exist, for example, similarity of KCs, dependency between KCs, and correspondence between questions and their KCs. To capture such relational structures for better addressing the KT problem, a recent trend is to explore the power of graph representation learning techniques such as Graph Neural Network (GNN). Table~\ref{tab:my_label} briefly summarizes several main graph-based knowledge tracing models and we discuss each of them separately.

\begin{table}[ht]
    \centering
       \begin{tabular}{ l |cc| l }\hline
   \multirow{2}{*}{\textbf{KT Model}}&\multicolumn{2}{c|}{\textbf{Graph}}&\multirow{2}{*}{\hspace{0.5cm}\textbf{Assumption}}\\\cline{2-3}
 & \textbf{Node} & \textbf{Edge}  &  \\ \hline
GKT~\cite{GKT19}  &KC & KC-KC relation   & One KC per question \\  
GIKT~\citep{GIKT20}  &{Question or KC}&Question-KC relation& Many KC per question\\
SKT~\cite{TongLHHCLM020} &KC&  KC-KC multiple relations  & One KC per question \\
\hline
    \end{tabular}
    \caption{A comparison of graph-based knowledge tracing models.}
    \label{tab:my_label}
\end{table}

\begin{itemize}
\item \textbf{Graph-based Knowledge Tracing (GKT)}

\emph{Graph-based Knowledge Tracing} (GKT), proposed by Nakagawa, Iwasawa, and Matsuo~\cite{GKT19}, attempted to incorporate a graph where nodes represent KCs and edges represent the dependency relation between KCs for a relational inductive bias. They reformulated the KT problem as a time series node-level classification problem and solved it using standard graph learning techniques such as message-passing GNNs~\cite{GNN_Model}.
Since such a graph is not explicitly given in KT tasks, the authors proposed two approaches to construct such graphs from a sequence of interactions by a student: (1) Statistics-based approach to construct a graph based
on statistics such as how many times one KC was answered after another KC was answered; 
(2) Learning-based approach to learn a graph with the performance optimization in an end-to-end manner.

\medskip
\item \textbf{Graph-based Interaction Knowledge Tracing (GIKT)}

\emph{Graph-based Interaction Knowledge Tracing} (GIKT), proposed by Yang et al.~\citep{GIKT20}, leverages the relation between questions and KCs, represented as a graph to learn useful embedding for answer prediction. Different from GKT which implicitly assumes that each question corresponds to one KC, GIKT assumes that one KC may be related to many questions and one question may correspond to more than one KC. Thus, GIKT can use a GNN to aggregate the embeddings of questions and KCs based on their relation in the graph, and sends the embedding of each question in a sequence of interactions to an RNN model to predict a student's answer for the next question.

\medskip
\item \textbf{Structure-based Knowledge Tracing (SKT)}

\emph{Structure-based Knowledge Tracing} (SKT) was proposed by Tong et al.~\cite{TongLHHCLM020} which aimed to capture multiple relations among KCs such as similarity relation and prerequisites relation. Similar to GKT, SKT also assumes that each question corresponds to one KC. However, instead of a single relation between KCs as captured in GKT, SKT exploits 
multiple relations between KCs. Further, SKT supports information propagation to jointly model the temporal and spatial effects when summarizing graph data. These two kinds of graph embeddings are combined at each time step and fed to a recurrent model to predict the correct answer by a student.

\end{itemize} 

\subsubsection{Text-Aware Knowledge Tracing Models}

Until now, the deep KT models that we have discussed mainly focused on a student's interactions with questions in a sequence to predict the probability of correctly answering the latest question by the student. Yet, they did not consider much about the textual features of questions themselves. Text-aware KT models are motivated by leveraging the textual features of questions to enhance the performance in tackling the KT tasks. 

\smallskip
\begin{itemize}
    \item \textbf{Exercise-Enhanced Recurrent Neural Network (EERNN)}

\emph{Exercise-Enhanced Recurrent Neural Network} (EERNN) was proposed by Su et al.~\cite{SuLLHYCDWH18}, which is a text-aware KT model to predict the probability of correctly answering a given question. The model uses a bi-directional LSTM module to extract the representation (i.e., a vector) of each question from the question's text and then trace a student's knowledge states by combining it with the representations of the previously answered questions using another LSTM module. Two variants of EERNN were developed: EERNNM, and EERNNA. The EERNNM variant assumes that a sequence of interactions satisfies the Markov property, i.e., the answer prediction for the next question only depends on the latest observed knowledge state; thus it only considers the last hidden state. The EERNNA variant considers all the previous knowledge states and combines them through an attention mechanism. 

\smallskip
Later, Yin et al.~\cite{YinLHCTW019} has further extended the work by Su et al.~\cite{SuLLHYCDWH18} through leveraging a pre-training task to learn question representations. The authors followed a masked language model (MLM) objective~\cite{word2vec} and showed that this pre-training step could further enhance the model's performance compared to the original model. 
\medskip

\item \textbf{Exercise-Aware Knowledge Tracing (EKT)}

\emph{Exercise-Aware Knowledge Tracing} (EKT)~\cite{LiuHYCXSH21} extends EERNN to incorporate the information of multiple KCs during answer prediction, where a student’s knowledge state is represented by a knowledge state matrix, rather than a knowledge state vector. Specifically, the model uses a memory network for quantifying how much each question can affect the mastery of a student on multiple KCs during a sequence of interactions by the student.
\end{itemize}\smallskip
In addition to the above text-aware KT models, other types of KT models such as \emph{Relation-Aware Self-Attention for Knowledge Tracing} (RKT)~\cite{pandey2020rkt} and \emph{Hierarchical Graph
Knowledge Tracing} (HGKT)~\cite{HGKT_20} also extract features from the textual information of questions for learning question representations in their models.

\subsubsection{Forgetting-Aware Knowledge Tracing Models}
 \label{subsec:forg}
   Learning psychological studies~\cite{Forgetting05,for15a,forlo16} showed that forgetting is an important aspect to consider for an accurate estimation of a student's knowledge state. This is because the knowledge mastery level of a student tends to decline with an exponential rate over time since the last practice of the relevant questions. From an experimental psychology perspective, Hermann Ebbinghaus~\cite{ebbinghaus2013memory,repeat_ebbinghouse_15} studied forces that affect memory retention, leading to formulate what is currently known as the \emph{learning curve theory}~\cite{learning_curve_theory}. Two effects that are reflecting these forces on memory retention are the \emph{forgetting} effect and the \emph{learning} effect.
 
 \medskip
  Modeling the forgetting effect is one of the major challenges that the KT literature has aimed at tackling. Traditional KT models have attempted to incorporate forgetting behavior by adding features such as the number of past trials or the lag
time from the previous interaction \cite{for16,pelanek2015modeling,qiu2011does,settles2016trainable}. In recent years, several deep learning KT models have been developed to take a student's forgetting behavior into consideration during tracing knowledge states.


\smallskip

\begin{itemize} 
\item \textbf{Deep Knowledge Tracing (DKT) + forgetting}

Nagatani et al.~\cite{wwwfOr19} proposed to extend the Deep Knowledge Tracing (DKT) model~\cite{piech2015deep} by adding sequence-related forgetting features. These features include: (1) the number of times a student answers questions with the same KC till the current point of time; (2) the time lapse since the last interaction on a question with the same KC; and, (3) the time lapse since the last interaction on a question regardless of its relating KC. The first feature  reflects the learning effect while the other two features reflect the forgetting effect. Different from the previous traditional KT models that use forgetting features only with regard to questions with the same KC \cite{for16,pelanek2015modeling,qiu2011does,settles2016trainable}, this work considers a student’s interactions in the whole sequence so as to model more complex forgetting behavior.

\medskip
\item\textbf{Knowledge Proficiency Tracing (KPT)}

\emph{Knowledge Proficiency Tracing} (KPT) was proposed by Chen et al.~\cite{for17}, which is a probabilistic matrix factorization model that leverages prior information for knowledge tracing. Specifically, two kinds of priors have been considered in this model: (1) \emph{question priors}: the model uses a Q-matrix which was marked by experts to depict the relationship between questions and KCs for generating question representations; and, (2) \emph{student priors}: the model captures the changes in a student’s knowledge state over time by jointly applying both learning curve and forgetting curve theories. The learning and forgetting factors are designed based on the assumption that a student’s current knowledge state is mainly influenced by two underlying reasons: (a) the more exercises she does, the higher level of related knowledge state she will get; and, (b) the more the time passes, the more knowledge she will forget. 
\smallskip

To further improve the predictive performance, an improved version of KPT, called \emph{Exercise-correlated Knowledge Proficiency Tracing}
(EKPT)~\cite{Zhenya_forget_20} was later developed, which incorporates the connectivity among questions over knowledge concepts into the probabilistic modeling.

\medskip

\item\textbf{HawkesKT}

 Inspired by the Hawkes process~\cite{hawkes1971spectra}, Wang et al.~\cite{TCEKT21} proposed \emph{HawkesKT}, a model that uses point process to adaptively model temporal cross-effects in KT. It assumes that the mastery of a KC by a student is not only affected by previous interactions on questions of the same KC, but also interactions on the other questions (\emph{cross-effects}). Further, the model assumes that cross effects caused by different previous interactions may also have different temporal evolutions on the mastery of different KCs. Although cross effects all decay with time, their decay rates differ from each other because some KCs may be easier to forget than the others.

\medskip
 
\item\textbf{Deep Graph Memory Network (DGMN)}

\emph{Deep Graph Memory Network} (DGMN)~\cite{abdelrahman2021deep} is a hybrid KT model that combines graph neural networks with memory for forgetting aware knowledge tracing. The model aims at modeling the forgetting behavior over a KC space, which has the advantage to capture indirect relationships between questions. DGMN builds a dynamic graph from a knowledge state memory to capture relationships across KCs. Given a sequence of interactions, DGMN uses an attention mechanism to associate questions to their relevant KCs. Then, it calculates forgetting features over the sequence, and fuses question embedding, KC graph embedding, and forgetting features using a gating mechanism. The gating output is used to predict the probability of answering the next question correctly.
\end{itemize}

\subsubsection{Discussion}
Deep learning KT models have demonstrated its great potential in solving the KT problem. Below, we discuss several key aspects that are crucial for being considered in their designs.\smallskip



\begin{itemize}
    \item \textbf{Knowledge state:} One fundamental assumption underlying each deep learning KT model is whether a knowledge state is considered over a single KC or multiple KCs. Accordingly, modeling a student's knowledge states based on the mastery level of KCs by the student is an important task in designing the deep learning KT models. Generally, from early works such as DKT which uses a hidden state to model knowledge states over a single KC, to later works by memory-augmented KT models (e.g., DKVMN and SKVMN) and by text-aware KT models (e.g., EKT) which use matrices to model knowledge states over multiple KCs, a trend in deep learning KT models is to develop a mechanism that is expressive for dynamically capturing knowledge state representations over complex KCs.
    \medskip
    
    \item \textbf{KC dependencies:} In a KT task, each question is assumed to associate with a single KC or multiple KCs, which is often provided as a prior knowledge such as Q-matrix. One main challenge faced by deep learning KT models is to discover the dependencies among different KCs, for example, one KC requires several other KCs as the prerequisite skills. To address this challenge, two lines of research have been explored in the KT literature, including: (1) using an attention mechanism to learn how questions are related to each other in terms of their required KCs; and, (2) using a graph-based learning model such as graph neural networks to learn the relationships between KCs or between questions according to their required KCs.
    
    \medskip
    \item \textbf{Feature augmentation:} To improve the model performance on KT tasks, additional features such as temporal features relating to forgetting behavior and textual features relating to question texts have been leveraged by a number of deep learning KT models in recent years. On one hand, augmenting additional features can usually lead to more accurate prediction on student learning performance; on the other hand, the augmentation of such additional features depends on their availability in databases, thus limiting their applicability within specific KT applications. 
\end{itemize}

  \begin{table}[t!]
   \centering
  \resizebox{1\textwidth}{!}{
  \begin{tabular}{l|c|c|c|c|cc}
    \toprule
\multirow{2}{*}{\textbf {KT Model}}&\multirow{2}{*}{\textbf{Learning Model}}&\multicolumn{2}{c|}{\textbf{Knowledge Component}}&\multirow{2}{*}{\textbf{Knowledge State}}&\multirow{2}{*}{\textbf{Forgetting}}\\\cline{3-4}&&\textbf{Single}&\textbf{Multiple}&&\\
     \midrule
    
\makecell{BKT~\citep{corbett1994knowledge}\\DBN~\cite{DBN}} &\makecell{HMM/BN\\DBN} &\makecell{$\checkmark$\\-}&\makecell{-\\$\checkmark$}&\makecell{Binary scalar\\Binary vector} &\makecell{$\times$\\$\times$}\\\hline

\makecell{IRT~\cite{IRT1951}\\HIRT~\cite{IRT2016back}\\TIRT~\cite{IRT2016back}\\LFA~\cite{AFM06}\\AFM~\cite{AFM08}\\PFA~\cite{FactorsAnalysis2009}\\KTM~\cite{KTM19}\\KTM-DLF~\cite{KTM-DLF020}}&\makecell{LR\\LR\\LR\\LR\\LR\\LR\\FM\\FM}&\makecell{$\checkmark$\\$\checkmark$\\$\checkmark$\\$\checkmark$\\$\checkmark$\\$\checkmark$\\-\\-}&\makecell{-\\-\\-\\-\\-\\-\\$\checkmark$\\$\checkmark$}&\makecell{Real-valued vector\\Real-valued vector\\Real-valued vector\\Real-valued vector\\Real-valued vector\\Real-valued vector\\Real-valued vector\\Real-valued vector}&\makecell{$\times$\\$\times$\\$\times$\\$\times$\\$\times$\\$\times$\\$\times$\\$\checkmark$}\\\hline

\makecell{DKT~\cite{piech2015deep}\\DKT+~\cite{YeungY18}\\DKT-DSC~\cite{DKT-DSC18}}&\makecell{RNN/LSTM\\RNN/LSTM\\RNN/LSTM}&\makecell{$\checkmark$\\$\checkmark$\\$\checkmark$}&\makecell{-\\-\\-}&\makecell{Hidden state (vector)\\Hidden state (vector)\\Hidden state (vector)}&\makecell{$\times$\\$\checkmark$\\$\times$}\\\hline

\makecell{DKVMN~\cite{zhang2017dynamic}\\SKVMN~\cite{SKVMN}}&\makecell{KVMN\\LSTM+KVMN}&\makecell{-\\-}&\makecell{$\checkmark$\\$\checkmark$}&\makecell{Key-value  memory (matrix)\\Key-value  memory (matrix)}&\makecell{$\times$\\$\times$}\\\hline

\makecell{SAKT~\cite{SAKT19}\\AKT~\cite{AKT20}\\SAINT~\cite{SAINT20}\\SAINT+~\cite{SAINT+21}\\RKT~\cite{pandey2020rkt}}&\makecell{FFN+MSA\\FFN+MSA\\FFN+ED+MSA\\FFN+ED+MSA\\FFN+MSA}&\makecell{-\\-\\-\\-\\-}&\makecell{$\checkmark$\\$\checkmark$\\$\checkmark$\\$\checkmark$\\$\checkmark$}&\makecell{Attentive embedding (matrix)\\Attentive embedding (vector)\\Attentive embedding (matrix)\\Attentive embedding (matrix)\\Attentive embedding (vector)}&\makecell{$\times$\\$\checkmark$\\$\times$\\$\times$\\$\times$}\\\hline

\makecell{GKT~\cite{GKT19}\\GIKT~\citep{GIKT20}\\SKT~\cite{TongLHHCLM020}}&\makecell{GNN\\GNN/RNN\\GNN}&\makecell{-\\-\\-}&\makecell{$\checkmark$\\$\checkmark$\\$\checkmark$}&\makecell{Vector\\Vector\\Vector}&\makecell{$\times$\\$\times$\\$\times$}\\\hline

\makecell{EERNN~\cite{SuLLHYCDWH18}\\EKT~\cite{LiuHYCXSH21}}&\makecell{RNN/LSTM\\AM/LSTM}&\makecell{$\checkmark$\\-}&\makecell{-\\$\checkmark$}&\makecell{Hidden state (vector)\\Attentive embedding (matrix)}&\makecell{$\times$\\$\times$}\\\hline

\makecell{DKT+forget~\cite{wwwfOr19}\\KPT~\cite{Zhenya_forget_20,for17}\\HawkesKT~\cite{TCEKT21}\\DGMN~\cite{abdelrahman2021deep}}&\makecell{RNN/LSTM\\FM\\FM\\GCN/KVMN}&\makecell{$\checkmark$\\-\\-\\-}&\makecell{-\\$\checkmark$\\$\checkmark$\\$\checkmark$}&\makecell{Hidden state (vector)\\ Real-valued vector\\ Real-valued vector\\Key-value  memory (matrix)}&\makecell{$\checkmark$\\$\checkmark$\\$\checkmark$\\$\checkmark$}\\
 \bottomrule
\end{tabular}}
  \caption{A summary of descriptive characteristics of main knowledge tracing models, where HMM is an abbreviation for Hidden Markov Mode~\cite{d2008more}, BN for Bayesian Network~\citep{villano1992probabilistic}, DBN for Dynamic Bayesian Network~\cite{DBN} , LR for Logistic Regression~\cite{logreg95}, FM for Factorization Machine~\cite{FM12,FM16}, GNN for Graph Neural Network~\cite{GNN_Model}, FFN for Feed Forward Network~\cite{Attention_17}, KVMN for Key-Value Memory Network~\citep{miller2016key}, ED for Encoder and Decoder model~\cite{Attention_17}, MSA for Multi-head Self-Attention mechanism or variants~\cite{Attention_17}) and AM for Attention Mechanism~\citep{NMT_15,Attention_17}. }
  \label{tbl:TKT}
\end{table}

\section{KNOWLEDGE TRACING DATASETS}
\label{Sec:DA}
This section presents an overview of the benchmark datasets used in the literature to support the evaluation of KT models. All publicly available datasets were downloaded, inspected and relevant information reported. Table~\ref{tbl:CD} lists the datasets and provides the general information such as student interactions, the number of questions, and data availability. More details about the datasets are presented below. 

\subsection{ASSISTments Datasets}
The ASSISTments datasets~\cite{feng2009addressing,pardos2014affective} contain longitudinal data collected from the free online tutoring ASSISTment platform\footnote{\url{https://www.assistments.org/}}. Table~\ref{tbl:ED} shows that the ASSISTments datasets are the most popular datasets used to benchmark KT models and the ones containing the most questions in total.\smallskip

These datasets are composed of grade school math exercises sampled from the \emph{Massachusetts Comprehensive Assessment System} (MCAS)\footnote{\url{https://www.doe.mass.edu/mcas/testitems.html}} containing different types of questions, such as multiple choice, text and open-ended questions. There are different versions of the ASSISTments datasets with data collected in different periods. Details about each version is presented below.\smallskip

\begin{itemize}
    \item \textbf{ASSISTments2009:} This dataset was collected during the school year $2009-2010$ and, when first released, contained a total of $525,535$ interactions (i.e., student responses to questions in the dataset), including duplicates, as discussed in~\cite{Going16}. The latest updated version of this dataset contains $346,860$ interactions given by $4,217$ students to a total of $26,688$ distinct questions and $123$ KCs. We note that only two thirds of the questions ($17,751$) are annotated with KCs. Questions without assigned KCs are annotated with `NA' (not available) or have no assigned value (`null') whereas all other questions are annotated with up to four KCs.
    
    Despite the popularity of this dataset, its original version is not reliable as discussed in~\cite{zhang2017dynamic} and the updated  `skill-builder'\footnote{\url{https://sites.google.com/site/assistmentsdata/home/assistment-2009-2010-data}} version is preferred as it fixes data modeling issues and removes duplicated records. It is also noteworthy to mention that results obtained with the different versions of this dataset (or with duplicated records) are often reported in the literature but they should not be directly compared to other approaches~\cite{IRT2016back}.
    \\
    \item \textbf{ASSISTments2012\footnote{\url{https://sites.google.com/site/assistmentsdata/home/2012-13-school-data-with-affect}}:} This is the largest version of the ASSISTments datasets consisting of data collected for one year (from Sept 2012 to Oct 2013). Despite the ASSISTments team reporting that the dataset contains approximately $10$ million `exercises', the available dataset consists of $179,999$ distinct questions answered by $46,674$ students resulting in $6,123,270$ interactions. We note that the vast majority ($126,908$) of the questions does not have any of the $265$ KCs associated with them. The lack of questions annotated with KCs may explain the overall lower performance of the KT models when applied to this dataset (see Table~\ref{tbl:CD}).
    \\
    \item \textbf{ASSISTments2015\footnote{\url{https://sites.google.com/site/assistmentsdata/home/2015-assistments-skill-builder-data}}:} This dataset contains $708,631$ student interactions with the ASSISTments platform in the year $2015$ produced by $19,917$ students answering $100$ distinct questions. The dataset contains only four attributes: (i) the questions' identifiers; (ii) the identifier of the student who answered the questions; (iii) an attribute indicating the correctness of the answer given by each student; and, (iv) a log attribute where a temporal sequence of the answers given by the students can be inferred. 
    
    Unlike previous versions of the ASSISTments datasets, no metadata and no KC are provided. Another difference between this and previous datasets is related to the average number of responses given to each question. This dataset has an average of $7,086.31$ answers per question whereas the $2009$ and $2012$ versions have $12.99$ and $34.01$, respectively. 
    \\
    \item \textbf{ASSISTment Challenge (ASSISTChall)\footnote{\url{https://sites.google.com/view/assistmentsdatamining}}:} Released in 2017, the full ASSISTment Challenge dataset contains data from $2004-2005$ and $2005-2006$ academic years. It contains $3,162$ distinct questions answered by $1,709$ students over $102$ KCs resulting in $942,816$ interactions. On average, this dataset has $298.17$ answers per question, placing it second in terms of the answer per question ratio (only lower than the ASSISTments2015 dataset). As part of a data mining competition, this dataset contains the most descriptive information among the ASSISTments datasets.
    
\end{itemize}

\subsection{STATICS2011 Dataset}
This dataset is available upon request and contains students' interactions with questions related to the Engineering Statics course\footnote{\url{https://pslcdatashop.web.cmu.edu/DatasetInfo?datasetId=507}} taught at the Carnegie Mellon University during Fall $2011$~\cite{STATICS2011}. 

\smallskip
The original dataset contains $361,092$ interactions, $335$ students and $1,224$ questions. In the KT literature, this dataset is often preprocessed~\cite{zhang2017dynamic} resulting in $1,223$ distinct questions answered by $333$ students over $85$ KCs. After preprocessing, the number of students' interactions is almost halved to $189,297$. This preprocessing was justified due to the large number of interactions without information on whether the questions have been correctly answered. The preprocessing considers the concatenation of the attributes `problem name' and `step name' and only the interactions with a valid first attempt. On average, this dataset has $568.45$ answers per question.

\subsection{Junyi Academy Dataset}
This dataset\footnote{\url{https://pslcdatashop.web.cmu.edu/DatasetInfo?datasetId=1198}} was collected between November 2010 and March 2015 from the Junyi Academy~\cite{Junyi}, an e-learning platform in Taiwan. The original dataset contains $25,925,992$ interactions, $247,606$ students, $722$ distinct questions, and $41$ KCs covering a number of topics in math. However, considering the same preprocessing step made in the previous datasets (i.e., students interactions with no hints given to help solve the questions and only students who have attempted each question once), the number of interactions drops to $21,571,469$ ($\sim17\%$), $220,441$ ($\sim11\%$) students, and $716$ ($<1\%$) distinct questions. Finally, on average, this dataset has $97.85$ answers per question. 

\smallskip
Although this dataset has been commonly used in the KT literature, the performance reported in some of the works cannot be directly compared. This is because these works use different subsets or preprocessing techniques~\cite{SKVMN, pandey2020rkt, TongLHHCLM020}. Further, note that an updated version of the Junyi dataset is available in Kaggle\footnote{\url{https://www.kaggle.com/junyiacademy/learning-activity-public-dataset-by-junyi-academy/tasks}} with data collected from August 2018 to August 2019. This dataset contains $11,468,379$ interactions, $25,649$ students, and $1,701$ distinct questions where no hints were given and students have attempted each question only once.

\subsection{Simulated-5 (Synthetic) Dataset}
This synthetic dataset was proposed by Piech et al.~\cite{piech2015deep}, which simulates virtual students to answer the same sequence of questions over a set of KCs in a controlled environment\footnote{\url{https://github.com/chrispiech/DeepKnowledgeTracing/tree/master/data/synthetic}}. The dataset is divided into two subsets, including training and testing. Each subset contains $50$ distinct questions associated with a single KC 
and a difficulty level. In total, each question is answered by $4,000$ virtual students resulting in $200,000$ interactions. The classic Item Response Theory~\cite{foster2017review} was used to create the interactions and simulate students learning over time~\cite{piech2015deep, YeungY18}. This dataset provides a standard format and does not require preprocessing steps such as removing duplicates or steps to infer a sequence of questions answered by a student, potentially allowing direct comparison between different KT models.

\subsection{KDDcup Dataset}
This dataset was presented at the KDDcup $2010$ Educational Data Mining challenge\footnote{\url{https://pslcdatashop.web.cmu.edu/KDDCup}}~\cite{kdd2010} and contains 13–14 year old students' responses to Algebra questions from $2005$ to $2007$ extracted from the intelligent tutoring system called ``The Cognitive Tutors'' developed by Carnegie Learning Inc. in the US. The dataset is split into three subsets described in what follows.
\\
\begin{itemize}
    \item \textbf{Algebra 2005-2006:} Considering both training and test data, collected between August $2005$ and June $2006$, this dataset contains $1,084$ distinct questions answered by $575$ students resulting in $813,661$ interactions. Unlike other datasets, each question is divided into sub-questions totaling $210,136$ sub-items. On average, there are $750.60$ answers per question. The total number of distinct KCs is $112$ and each sub-question is associated with one or more KCs. As in previous datasets, the total number of interactions drops to $57.8\%$, the number of students roughly remains the same ($574$) and the number of sub-items is reduced to $130,256$ if the data is preprocessed removing sub-items with no KCs assigned, interactions where students used any hints and questions with two or more attempts.
    \\
    
    \item \textbf{Algebra 2006-2007:} This subset contains $2,289,726$ interactions generated by $1,840$ students answering $90,831$ distinct questions with a total of $577,119$ sub-items classified by one or more of the $523$ KCs. On average, there are $49.36$ answers per question. Similarly to the Algebra 2005-2006 dataset, the number of interactions drops to $1,567,072$, the number of students falls to $1,813$ and the number of questions and sub-items reduce to $89,877$ and $470,187$, respectively, after the data preprocessing step. After inspecting this subset, we found that the timestamps provided are incorrect. This may affect the prediction of KT approaches and may explain the low adoption of this subset in comparison to the others. 
    \\
    
    \item \textbf{Bridge to Algebra:} This subset contains $3,686,871$ interactions collected between November 2006 and June 2007 generated by $1,146$ students, $19,258$ distinct questions divided into $207,790$ sub-items classified by one or more of the $493$ KCs. After preprocessing, the total number of interactions is almost halved ($1,721,987$), the number of questions drops ($18,715$) slightly whereas the number of items is reduced to $123,778$ ($\sim40\%$) and the number of students remains almost the same ($1,145$).
\end{itemize}

\subsection{EdNet Dataset}
EdNet\footnote{\url{https://github.com/riiid/ednet}} is a hierarchical dataset composed of four subsets identified by the ids: `KT1', `KT2', `KT3' and `KT4', each containing different types of student activities. `KT1', for example, contains question-response pairs similar to other datasets. The main difference is that some questions in this dataset are organized in bundles (a set of questions that must be completed altogether). This dataset contains over $95,293,926$ interactions, $13,169$ questions, $784,309$ students, and $188$ KCs. 

\smallskip
Unlike the other datasets, `KT2' contains the actions of the users during question-solving activities. For example, it records the final submission and student decision-making (alternating choices) before submitting the final answer. This subset has $56,360,602$ interactions and $297,444$ students. Besides the actions in `KT1' and `KT2', `KT3' subset includes information about how students interact with learning activities to answer a question (e.g., watch a lecture). This subset contains more KCs ($293$), interactions ($89,270,654$) and students ($297,915$). Finally, `KT4' is the most complete subset containing every action recorded by the EdNet system, including students' purchases (e.g., course purchases). This subset contains $131,441,538$ interactions in total.

\smallskip
Overall, the EdNet dataset series incrementally provides information into student activities and behaviors. The dataset was collected over two years from the intelligent online tutoring platform named \emph{Riid TUTOR}\footnote{\url{https://company.riiid.co/en/product}} dedicated to practicing English for international communication (TOEIC) assessment~\cite{EDNET2020ednet} in South Korea. The variety of recorded behaviors and the large size of data points are unique aspects of this dataset.

 \begin{table*}

  \begin{tabular}{l|ccccc}
    \toprule
     \multirow{2}{*}{\textbf{Dataset}}&\multirow{2}{*}{\textbf{\#Questions}} &\multirow{2}{*}{\textbf{\#Students}}&\multirow{2}{*}{\textbf{\#Interactions}}&\multirow{2}{*}{\textbf{\#KCs}}&\multirow{1}{*}{\textbf{Public}}\\
     &&&& &\multirow{1}{*}{\textbf{available}}\\
     
      \midrule
       ASSISTments2009&$26,688$&$4,217$&$346,860$&$123$&Yes\\
       ASSISTments2012&$179,999$&$46,674$&$6,123,270$&$265$&Yes\\
       ASSISTments2015&$100$&$19,917$&$ 708,631$& - &Yes\\
       ASSISTChall&$3,162$&$1,709$&$942,816$&$102$&Yes\tabularnewline
        \midrule
       STATICS2011&$ 1,224$&$335$&$361,092$&$85$&No\tabularnewline
        \midrule
        Junyi Academy&$722$&$247,606$&$25,925,992$&$41$&Yes\tabularnewline
        \midrule
        Simulated-5 (Synthetic)&$50$&$4,000$&$ 200,000$&$5$&Yes\tabularnewline
        \midrule
        Algebra 2005-2006&$1,084$&$575$&$813,661$&$112$&Yes\\
        Algebra 2006-2007&$90,831$&$1,840$&$2,289,726$&$523$&Yes\\
        Bridge to Algebra &$19,258$&$1,146$&$3,686,871$&$493$&Yes\tabularnewline
        \midrule
        EdNet-KT1&$13,169$&$784,309$&$ 95,293,926$ &$188$&Yes \\
        EdNet-KT2&$13,169$&$297,444$& $56,360,602$&$188$&Yes\\
        EdNet-KT3&$13,169$&$297,915$&$ 89,270,654$&$293$&Yes\\
        EdNet-KT4&$13,169$&$297,915$&$ 131,441,538$&$293$&Yes\\
   \bottomrule
\end{tabular}
\caption{Dataset Statistics.}
  \label{tbl:CD}
\end{table*}

\subsection{Considerations}
Table~\ref{tbl:ED} presents the results obtained by several KT models using the aforementioned datasets. The results are reported using the AUC-ROC curve --- a traditional performance measurement for binary classification models. The probabilistic receiver operating characteristic (ROC) curve plots the true positive rate (TPR) against the false positive rate (FPR) while the area under the curve (AUC) reports how well a KT model can distinguish between correct and incorrect answers. The AUC-ROC ranges from 0 to 1, where 0.5 indicates an uninformative classifier (random guesses) and 1 a perfect classifier.

It is important to note that the results cannot be directly compared if experimental settings are not standardized. As discussed in~\cite{IRT2016back}, the results obtained without removing duplicate interactions (preprocessing steps) may be inaccurate and appear elevated. The existence of duplicates is also acknowledged in the KDDcup dataset\footnote{\url{https://pslcdatashop.web.cmu.edu/KDDCup/FAQ/}}. We reported in the previous sections the raw number of interactions, students, KCs, etc., and whenever different, the total after removing duplicates and discarding noise data.

\begin{landscape}
  \begin{table}[ht!]
   \centering
 \begin{adjustbox}{width=0.9\linewidth}
 \begin{tabular}{|l|c|c|c|c|c|c|c|c|c|c|c|}
    \toprule
     \multirow{3}{*}{\textbf{Model}}& \multirow{3}{*}{\textbf{Data Split Ratio}}&\multicolumn{10}{c|}{\textbf{Dataset}}\\\cline{3-12}
     &&\multicolumn{4}{|c|}{ \textbf{ASSISTments}}&\multirow{2}{*}{\textbf{STATICS}}&\multirow{2}{*}{ \textbf{Junyi}}&\multirow{2}{*}{\textbf{Synthetic}}&\multicolumn{2}{|c|}{\textbf{KDDcup}}&\multirow{2}{*}{\textbf{EdNet}} \\
     \cline{3-6} \cline{10-11}
    &\textbf{(Train-Test)}&2009 & 2012 & 2015 &Chall&&&&Algebra 2005-2006&Bridge 2006-2007&
     \\ 
    \midrule 
     BKT~\citep{corbett1994knowledge}&$80\%$-$20\%$&\makecell{$0.651$~\cite{DKT-DSC18}\\$0.67$~\cite{piech2015deep}\\$0.6571$~\citep{GIKT20}}&\makecell{$0.623$~\cite{DKT-DSC18}\\$0.6204$~\citep{GIKT20}}&\makecell{$0.611$~\cite{DKT-DSC18}\\$0.678$~\cite{TongLHHCLM020}}&-&-&\makecell{$0.68$~\cite{piech2015deep}\\$0.831$~\cite{TongLHHCLM020}}&$0.54$~\cite{piech2015deep}&$0.642$~\cite{DKT-DSC18}&-&$0.6027$~\citep{GIKT20}\\\hline
     
     IRT~\cite{IRT1951}&$80\%$-$20\%$&\makecell{$0.7651$~\cite{IRT2016back}\\$0.6908$~\cite{KTM19}\\$0.751$~\cite{DKT-DSC18}\\$0.5869$~\cite{TCEKT21}}&\makecell{$0.702$~\cite{KTM-DLF020}\\$0.743$~\cite{DKT-DSC18}\\$0.6340$~\cite{TCEKT21}}&$0.672$~\cite{DKT-DSC18}&-&-&-&-&\makecell{$0.771$~\cite{KTM-DLF020}\\$0.806$~\cite{DKT-DSC18}}&\makecell{$0.747$~\cite{KTM-DLF020}\\$0.8542$~\cite{IRT2016back}}&-\\\hline
     
     HIRT~\cite{IRT2016back}&$80\%$-$20\%$&$0.774$~\cite{IRT2016back}&-&-&-&-&-&-&-&$0.8597$~\cite{IRT2016back}&-\\\hline
     
     TIRT~\cite{IRT2016back}&$80\%$-$20\%$&$0.7653$~\cite{IRT2016back}&-&-&-&-&-&-&-&$0.8542$~\cite{IRT2016back}&-\\\hline
     
     AFM~\cite{AFM08}&$80\%$-$20\%$&$0.6163$~\cite{KTM19}&$0.610$~\cite{KTM-DLF020}&-&-&-&-&-&$0.707$~\cite{KTM-DLF020}&$0.706$~\cite{KTM-DLF020}&-\\\hline
     
     PFA~\cite{FactorsAnalysis2009}&$80\%$-$20\%$&\makecell{$0.6849$~\cite{KTM19}\\$0.703$~\cite{DKT-DSC18}}&\makecell{$0.670$~\cite{DKT-DSC18}\\$0.669$~\cite{KTM-DLF020}}&$0.689$~\cite{DKT-DSC18}&-&-&-&-&\makecell{$0.760$~\cite{DKT-DSC18}\\$0.744$~\cite{KTM-DLF020}}&$0.746$~\cite{KTM-DLF020}&-\\\hline
     
     KTM~\cite{KTM19}&$80\%$-$20\%$&\makecell{$0.8186$~\cite{KTM19}\\$0.7169$~\citep{GIKT20}\\$0.7425$~\cite{TCEKT21}}&\makecell{$0.7535$~\cite{TCEKT21}\\$0.6788$~\citep{GIKT20}}&-&-&-&-&-&-&-&$0.6888$~\citep{GIKT20}\\\hline
     
     KTM-DLF~\cite{KTM-DLF020}&$80\%$-$20\%$&&$0.756$~\cite{KTM-DLF020}&-&-&-&-&-&$0.837$~\cite{KTM-DLF020}&$0.812$~\cite{KTM-DLF020}&-\\\hline
     
     DKT~\cite{piech2015deep}&$80\%$-$20\%$&\makecell{$0.7429$~\cite{IRT2016back}\\$0.86$~\cite{piech2015deep}\\$0.8212$~\cite{YeungY18}\\$0.721$~\cite{DKT-DSC18}\\$0.820$~\cite{SAKT19}\\$0.817$~\cite{AKT20}\\$0.709$~\cite{GKT19}\\$0.7561$~\citep{GIKT20}\\$0.7515$~\cite{TCEKT21}}&\makecell{$0.713$~\cite{DKT-DSC18}\\$0.712$~\cite{pandey2020rkt}\\$0.7286$~\citep{GIKT20}\\$0.7235$~\cite{wwwfOr19}\\$0.7308$~\cite{TCEKT21}}&\makecell{$0.707$~\cite{DKT-DSC18}\\$0.731$~\cite{AKT20}\\$0.736$~\cite{SAKT19}\\$0.7365$~\cite{YeungY18}\\$0.727$~\cite{TongLHHCLM020}}&\makecell{$0.7263$~\cite{AKT20}\\$0.734$~\cite{SAKT19}\\$0.7343$~\cite{YeungY18}}&\makecell{$0.8233$~\cite{AKT20}\\$0.815$~\cite{SAKT19}\\$0.8159$~\cite{YeungY18}}&\makecell{$0.814$~\cite{pandey2020rkt}\\$0.85$~\cite{piech2015deep}\\$0.847$~\cite{TongLHHCLM020}}&\makecell{$0.8255$~\cite{YeungY18}\\$0.823$~\cite{SAKT19}\\$0.75$~\cite{piech2015deep}}&$0.784$~\cite{DKT-DSC18}&\makecell{$0.751$~\cite{GKT19}\\$0.8901$~\cite{IRT2016back}}&\makecell{$0.6822$~\citep{GIKT20}\\$0.7638$~\cite{SAINT+21}}\\\hline
     
     DKT+~\cite{YeungY18}&$80\%$-$20\%$&\makecell{$0.8227$~\cite{YeungY18}\\$0.8024$~\cite{AKT20}\\$0.822$~\cite{SAKT19}}&-&\makecell{$0.7371$~\cite{YeungY18}\\ $0.728$~\cite{TongLHHCLM020}\\$0.737$~\cite{SAKT19}\\$0.7313$~\cite{AKT20}}&\makecell{$0.7343$~\cite{YeungY18}\\$0.728$~\cite{SAKT19}\\$0.7124$~\cite{AKT20}}&\makecell{$0.8349$~\cite{YeungY18}\\$0.835$~\cite{SAKT19}\\$0.8301$~\cite{AKT20}}&$0.889$~\cite{TongLHHCLM020}&\makecell{$0.8264$~\cite{YeungY18}\\$0.824$~\cite{SAKT19}}&-&-&-\\\hline
     
     DKT-DSC~\cite{DKT-DSC18}&$80\%$-$20\%$&$0.735$~\cite{DKT-DSC18}&$0.721$~\cite{DKT-DSC18}&$0.716$~\cite{DKT-DSC18}&$0.430$~\cite{DKT-DSC18}&-&-&-&$0.792$~\cite{DKT-DSC18}&-&-\\\hline
     
     DKVMN~\cite{zhang2017dynamic}&$70\%$-$30\%$&$0.8157$~\cite{SKVMN}~\cite{zhang2017dynamic}&-&$0.7268$~\cite{SKVMN}~\cite{zhang2017dynamic}&-&$0.8284$~\cite{SKVMN}~\cite{zhang2017dynamic}&$0.8027$~\cite{SKVMN}&$0.8273$~\cite{SKVMN}~\cite{zhang2017dynamic}&-&-&-\\\hline
     
     SKVMN~\cite{SKVMN}&$70\%$-$30\%$&$0.8363$~\cite{SKVMN}&-&$0.7484$~\cite{SKVMN}&-&$ 0.8485$~\cite{SKVMN}&$0.8267$~\cite{SKVMN}&$0.840$~\cite{SKVMN}&-&-&-\\\hline
     
     SAKT~\cite{SAKT19}&$80\%$-$20\%$&\makecell{$0.848$~\cite{SAKT19}\\$0.752$~\cite{AKT20}\\$0.6860$~\cite{TCEKT21}}&\makecell{$0.735$~\cite{pandey2020rkt}\\$0.6906$~\cite{TCEKT21}}&\makecell{$0.854$~\cite{SAKT19}\\$0.7212$~\cite{AKT20}}&\makecell{$0.734$~\cite{SAKT19}\\$0.6569$~\cite{AKT20}}&\makecell{$0.853$~\cite{SAKT19}\\$0.8029$~\cite{AKT20}}&$0.834$~\cite{pandey2020rkt}&$0.832$~\cite{SAKT19}&-&-&\makecell{$0.7671$~\cite{SAINT20}\\$0.7663$~\cite{SAINT+21}}\\\hline
     
     AKT~\cite{AKT20}&$80\%$-$20\%$&\makecell{$0.8346$~\cite{AKT20}\\$0.7474$~\cite{TCEKT21}}&$0.7555$~\cite{TCEKT21}&$0.7828$~\cite{AKT20}&$0.7702$~\cite{AKT20}&-&-&-&-&-&-\\\hline
     
     SAINT~\cite{SAINT20}&$80\%$-$20\%$&-&-&-&-&-&-&-&-&-&\makecell{$0.7811$~\cite{SAINT20}\\$0.7816$~\cite{SAINT+21}}\\\hline
     
     SAINT+~\cite{SAINT+21}&$80\%$-$20\%$&-&-&-&-&-&-&-&-&-&$0.7914$~\cite{SAINT+21}\\\hline
     
     RKT~\cite{pandey2020rkt}&$80\%$-$20\%$&-&$0.793$~\cite{pandey2020rkt}&-&-&-&$0.860$~\cite{pandey2020rkt}&-&-&-&-\\\hline
     
     GKT~\cite{GKT19}&$80\%$-$20\%$&$0.723$~\cite{GKT19}&$0.735$~\cite{TongLHHCLM020}&-&-&-&$0.893$~\cite{TongLHHCLM020}&-&-&$0.769$~\cite{GKT19}&-\\\hline
     
     GIKT~\citep{GIKT20}&$80\%$-$20\%$&$0.7896$~\citep{GIKT20}&$0.7754$~\citep{GIKT20}&-&-&-&-&-&-&-&$0.7523$~\citep{GIKT20}\\\hline
     
     SKT~\cite{TongLHHCLM020}&$80\%$-$20\%$&-&-&$0.746$~\cite{TongLHHCLM020}&-&-&$0.908$~\cite{TongLHHCLM020}&-&-&-&-\\\hline
    
    DKT+forget~\cite{wwwfOr19}&$80\%$-$20\%$&$0.754$~\cite{TCEKT21}&\makecell{$0.722$~\cite{pandey2020rkt}\\$0.7309$~\cite{wwwfOr19}\\$0.7462$~\cite{TCEKT21}}&-&-&-&$0.840$~\cite{pandey2020rkt}&-&-&-&-\\\hline
     
     HawkesKT~\cite{TCEKT21}&$80\%$-$20\%$&$0.763$&$0.768$&-&-&-&-&-&-&-&-\\\hline
     DGMN~\cite{abdelrahman2021deep}&$70\%$-$30\%$&$86.1$~\cite{abdelrahman2021deep}&-&-&-&$ 86.4$~\cite{abdelrahman2021deep}&-&$85.9$~\cite{abdelrahman2021deep}&$83.4$~\cite{abdelrahman2021deep}&-&-\\
  \bottomrule
\end{tabular}
\end{adjustbox}  \caption{AUC results reported by different KT models.}
  \label{tbl:ED}
\end{table}
\end{landscape}

The lack of accurate and descriptive information about each of the attributes in the datasets also hinders experiments. An example is in the ASSISTments datasets where terminology used is confusing\footnote{\url{https://sites.google.com/site/assistmentsdata/home/2012-13-school-data-with-affect}} or lacking\footnote{\url{https://sites.google.com/site/assistmentsdata/home/2015-assistments-skill-builder-data}}. This may explain the different AUC values reported to the same approach and dataset pairs presented in Table~\ref{tbl:ED}.

The sequence of students' interactions is also an important component of a KT task that is affected by the quality of the datasets. Due to noise in the data (e.g., incorrect timestamps, null values, etc.), the sequence of interactions may not be correctly extracted from the datasets which may accordingly impact the performance of the proposed KT models. For example, after inspecting the datasets, we identified timestamp errors in the Algebra $2006-2007$, which may justify its low adoption in comparison to the other two Algebra datasets.

The fact that the benchmark datasets are often used for multiple tasks other than the KT problem hinders the correct use and interpretation of the datasets in the KT context. The works in \cite{DKT-DSC18, YeungY18, GIKT20, zhang2017dynamic, SAKT19, AKT20, TCEKT21}, for example, report different numbers of knowledge components for the same datasets and the work in \cite{AKT20} discusses how different experimental settings (association between KC and questions) may impact the reported results. The data model and the file format chosen to represent the datasets may also contribute to misinterpretation of the data. For example, the data is often extracted from relational databases and stored in comma-separated values (CSV), compressing information in a single attribute using non-standard characters, e.g., the KDDcup dataset uses double tilde characters to assign multiple KCs to a question. Given the hierarchical and relational nature of the data, XML and JSON are more suitable and explicit formats.

Another critical aspect is the lack of more diverse datasets. The existing public datasets are often from a specific domain (mainly math) and a specific region (e.g., the ASSISTment datasets from the US, and EdNet, the most recent publicly available dataset, from South Korea). Most of the datasets do not provide demographic information, and therefore gender-based, or other similar predictions cannot be performed.

Finally, the benchmark datasets have been updated over the years, and the version used by the KT models is not readily identifiable, which can compromise their direct comparison. A version control mechanism for the publicly available datasets would help keep track of every change made to a dataset over time and allows for consistent and comparable results.

\section{KNOWLEDGE TRACING APPLICATIONS}
\label{sec:kt_apps}

This section explores possible application areas that can benefit from KT models. We broadly divide these application areas into the following four categories: (i) Recommender Systems; (ii) Learning Provision and Quality Assurance; (iii) Interactive Learning; and, (iv) Learning to Teach.

\subsection{Recommender Systems}
An application area of KT that comes directly to mind is online education systems where the main objective is to provide effective learning experience for their students. Tracing the knowledge state of a student would facilitate to tailor students' learning experience according to their capabilities and skills. This can be achieved through recommending learning materials (e.g., lectures, labs, and/or exercises) based on the learnt knowledge state of a student. Thus, the aim of such online education systems is twofold: (1) to estimate the knowledge state of a student using a KT model; and, (2) to recommend learning materials conditioned on the knowledge state using a recommendation model~\cite{BOBADILLA2013109}. Below are some examples of applications that have been recently studied. A graph-based recommendation method has been proposed by Chanaa and Faddouli~\cite{lmr_chanaa} to aid an educational instructor to segment students into groups based on their knowledge states and recommend the most appropriate kinds of exercises for each group. More specifically, the instructor first selects a specific knowledge component (KC), and then, the model constructs a dynamic knowledge graph based on historical practice information that includes student knowledge vectors as nodes and uses edges for representing their mastery level similarities around the selected KC. Such a graph is clustered into node groups and for each group a shared embedding is constructed using GNNs to get the final recommendations. Cai et al.~\cite{lmr_cai} followed an interactive recommendation approach in which a reinforcement learning recommender agent selects learning materials to recommend based on a reward signal calculated from the progress in a knowledge state estimated by a KT model. Huang et al.~\cite{lmr_huang} proposed an interactive educational video recommender model that follows a multi-objective setup of rewards. The authors designed three reward functions to reflect three main aspects in online education systems including a reviewing reward function for recommending videos about KCs that a student did not perform well previously, a smoothing reward function for recommending videos with a gradual difficulty to understand, and an engagement reward function for recommending videos about KCs a student started to master recently. The recommender agent followed a reinforcement learning design with a state estimated by a DKT variant model, an action for the video id to recommend, and a combined reward by using a weighted sum of the three reward functions.

\subsection{Learning Provision and Quality Assurance}
Another potential application area is to provision a learning curriculum (e.g., an ordered list of topics to study) for a specific subject based on the knowledge state of a student. One direction~\cite{lcp_corbett,lcp_schodde} in this application area follows a semi-automated approach in which an instructor would conduct simulations using a KT model trained on the historical exercise records of students to identify a suitable curriculum of learning materials for a course to maximize the knowledge gain of students. Another direction~\cite{Pardos_13} uses KT models to assess the effectiveness of a course structure in achieving its targeted objectives by assessing the impact of each module (i.e., a  collection of learning materials) on the knowledge growth of students. Recently, deep KT models have been adopted in this direction to provide quality assurance for the course design~\cite{Liu_kdd_19}. The authors used a DKT model~\cite{piech2015deep} to trace the progress in the knowledge state of a student after taking a specific course module, and then, an actor-critic reinforcement learning agent~\cite{actor_critic_rl} considers the knowledge state across different course modules and their predefined relationships (i.e., prerequisites) and takes an action to select the next module to work on for the student to maximize their knowledge gain in achieving course objectives. Following this approach, the structure of a course could adapt dynamically to a student's needs and skills instead of having one fixed structure that does not fit all students.

\subsection{Interactive Learning}
Interactive education aims at making the learning process more exciting and engaging through delivering knowledge components into a gaming shell. Cognitive studies~\cite{edu_gaming_cog1} showed that students can easily gain new knowledge components and be able to spend more time on learning if the learning materials were delivered in an engaging manner such as gaming. This is due to the nature of the human mind that was mainly designed for learning by real-world practices that usually come as a form of an interactive experience (i.e., state, action, and rewards). Thus, an educational game can provide a similar experience that could be more aligned with our natural learning capabilities than the conventional educational ways (e.g., textbooks, lectures, etc.) that lack engaging interaction. Long and Aleven~\citep{edu_long_17} evaluated the effect of educational games in comparison to conventional online tutoring systems through a study that involved two groups of students: one group was learning on an educational game for math equation solving and the other group was learning using a non-interactive system that presents math concepts in a conventional manner through demonstrative examples and exercises. The study found that the group using the educational game was more excited and engaged to continue the learning procedure in comparison to the other group. Another study~\cite{edu_gaming_mob} focused on the effect of mobile educational games on the learning progress for elementary school kids. The authors divided the student into two groups: one allowed to use mobile educational games to review and practice math concepts taught at school and another group practiced the math concepts using conventional text exercises. The study concluded that students with access to the mobile educational games were performing better in retaining the math concepts in comparison to the other group. These findings about the effectiveness of educational games demonstrate the great potential of education games as an promising application area for KT models. 

For an educational game to be more effective, it has to assess the knowledge progress of a player and adjust the gaming experience accordingly. This might include adjusting the difficulty of challenges, opening new part(s) in a game, or adjusting the competency of a computer opponent. Kantharaju et al.~\cite{Kantharaju_18} proposed a KT model that could detect when a player attempts a specific skill in an educational game and quantify their knowledge state across different states, so the game experience could be focused on challenging skills for each player. Cui et al.~\cite{cui2019analyzing} used the BKT~\cite{BKT1992probabilistic} model to trace the knowledge state of fifth-grade elementary school students in Canada during a science gaming assessment. The authors were not only able to effectively predict the final score of a student based on their partial observations from the game assessment, but also identify pitfalls in the game design and assumptions that tend not to work as expected by the designer in terms of assessing dedicated skills.

\subsection{Learning to Teach}
Going beyond the conventional assumption of a human student in a KT setting opens the door to a wide range of application areas. Virtual students, such as intelligent agents which adopt a reinforcement learning setup or machine learning models, can be treated in a similar way as real-life students who are in need to learn a set of skills in different machine learning tasks. For example, this can be a deep neural network model that needs to master a skill of classifying different class labels (e.g., cats, dogs, furniture, etc.) in an image classification task or a reinforcement learning agent aiming at mastering different skills in an Atari game. Curriculum learning (CL)~\cite{bengio_cl} aims at learning a curriculum of tasks to enable a student agent from mastering a set of skills. A CL policy would imply a statistical distribution on learning tasks that gradually drive the student agent towards convergence. Another relevant paradigm is machine teaching (MT)~\citep{Zhu_2015} which aims at minimizing the teaching cost represented by the size of training samples drawn from training data in a machine learning scenario. In MT, there are two models being included: the teacher model and the student model. The former targets to sample training data for the latter to learn an optimal parameter set $\theta^*$ the minimizes the loss function in the task. Learning to teach (L2T)~\cite{l2t_1,l2t_2} targeted customizing the learning process for a student agent/model through optimizing three main aspects including training data sampling, neural architecture design, and loss function design. In L2T, a teacher agent follows a reinforcement learning approach to optimize a teaching policy that handles one or more of the three aspects previously mentioned. It can be observed that a shared characteristic across these different attempts to enhance the conventional machine learning procedure is the need to trace the knowledge state of a student model. Thus, there is a significant potential for KT models to contribute in this application area by tracing the knowledge state of a student model during training procedure. The output of the KT model would form the input/state of a teacher model that aims at customizing the training procedure to speedup the student model's convergence. Knowledge Augmented Data Teaching (KADT)~\cite{abdelrahman2021learning} aims at improving a data teaching strategy of a student ML model by tracking its knowledge progress across multiple knowledge components in a learning task. The KADT method includes a knowledge tracking model to dynamically capture the knowledge progression of the student model in terms of latent knowledge components. The authors develop an attention-pooling mechanism to extract knowledge representations of the student model with respect to class labels, which enables the development of a data-teaching strategy on significant training samples. The authors evaluated the performance of the KADT method on four different machine learning tasks including knowledge tracing, sentiment analysis, movie recommendation, and image classification. Results compared to the state-of-the-art machine teaching methods have been empirically proven that KADT consistently outperforms the others in all tasks.

\section{KNOWLEDGE TRACING FUTURE RESEARCH DIRECTIONS}
\label{sec:future}

Despite the promising results achieved by current state-of-the-art KT models, limitations and gaps of current approaches and available datasets open up several opportunities for future research. A numbered list of such research opportunities is presented below.

\medskip
\noindent\textbf{Multimodal and informative representation learning \& datasets}. The choice of data representation directly impacts the performance of any machine learning model \cite{10.1109/TPAMI.2013.50}. KT models tend to learn embedding representations for questions and KCs from abstract formats such as one-hot encoding; however, some data in the description of a question such as images and mathematical equations that can lead to more informative embedding representations, are overlooked, either by the proposed models or by the available datasets. This opens up research directions (RD) through the following questions: 
\begin{itemize}
    \item[($RD_1$)] What information/satellite data can be used to improve the performance of KT models?
    \item[($RD_2$)] How to represent such data for the KT tasks? 
    \item[($RD_3$)] How to create a dataset for the KT tasks that enables a more informative embedding representation learning?
\end{itemize}

The Exercise-aware Knowledge Tracing (EKT) approach proposed by Liu et al. \cite{LiuHYCXSH21} is a recent attempt to learn richer embedding representations, taking into account the textual context and the relationships between questions. However, despite the previous efforts, the representations of multimodal and domain-specific data such as mathematical equations and code snippets remain mostly unexplored in the literature resulting in low-informative representation learning for KT models. Fusing signals from multiple feature spaces may enable better representation learning in addition to mitigating noise in data~\cite{mml_19}.

To solve ($RD_1$) and ($RD_2$), a new range of benchmark datasets for KT must be created containing contextual information rather than only encoded data with ids (as previously seen in Section \ref{Sec:DA}). Datasets spanning various knowledge domains and from different cultures and educational levels are also needed, as the performance of KT models can be affected when applied to other demographic and educational contexts. This leads us to the next research opportunity, Self-Supervised Learning in Knowledge Tracing.
 
\medskip 
\noindent\textbf{Self-supervised learning in knowledge tracing}. Although supervised learning has led to advances in different areas, it still has a major drawback: the need for large, high-quality labeled data for training. Self-supervised learning (SSL)~\cite{Zhai_2019_ICCV, Misra_2020_CVPR}, on the other hand, has proved to be effective in several areas (e.g., natural language processing~\cite{BERT_19} and computer vision~\cite{SSL_CV}) learning from unlabeled data. SSL often adopts similarity ranking loss functions (e.g., contrastive loss~\cite{Wang_2021_CVPR}) in a process called \emph{pre-training} or \emph{pretext task}~\cite{Misra_2020_CVPR} to automatically generate labels. Such pre-trained models can thus be transferred to a downstream task to train in a supervised learning manner with a limited amount of labeled data. Along with SSL, therefore, further contributions to the KT field can be made, for example, by ($RD_4$) creating pre-trained models (e.g., using existing pre-trained language and computer vision models) to generate informative representations for KT; and, ($RD_5$) investigating how it can mitigate limited training of students' activities in cases of a cold-start scenario or skewed participation data, e.g., when only a small number of students contribute to most activities.

 \medskip
\noindent\textbf{Interactive knowledge tracing}. Most KT models adopt a passive approach of observing question answering response history to estimate students' knowledge states; however, interactive methods, driven by question answering response behavior, to better understand the dynamics of students' knowledge states are still unexplored. Interactive methods are particularly useful in cold start scenarios where an interactive approach can reveal students' knowledge states by directly asking questions related to different KCs. Thus, another potential future work is to ($RD_6$) develop optimized question sampling policies to enhance the performance of KT models in, but not limited to, cold start scenarios. Reinforcement Learning (RL) ~\cite{sutton2018reinforcement} approaches are a possible natural choice given its maximal rewarding scheme.
 
Last but not least, considering that Knowledge Tracking involves human knowledge and learning, transparency in the internal logic and the results obtained by KT models would benefit educational stakeholders and processes. This leads us to investigate on \textbf{eXplainable Artificial Intelligence} (XAI) approaches, as seen in other research fields such as \cite{exp_ai_fair_20, exp_ai_ijaci20}. Potential research avenues includes ($RD_7$) the development of techniques and methods to understand and explain the prediction process in KT models; and, ($RD_8$) how algorithmic decisions make impacts on learning processes, course design, instructor performance, the quality of learning materials, and student engagement. Promising research to explain deep learning models has been carried out using knowledge distillation ~\cite{Haselhoff_2021_CVPR} to understand and explain predictions in other models.
\section{CONCLUSION}
\label{sec:conclusion}

In this work, we presented a comprehensive research monograph for 
knowledge tracing. To cover the related fundamental concepts, we covered early attempts for knowledge tracing in chronological order and illustrated related background concepts. We built upon the historical Knowledge Tracing landscape to layout a categorization for the relevant literature based on shared theoretical and algorithmic aspects. 

As deep learning is considered the current prominent toolkit for the majority of the state-of-the-art KT approaches, we deeply reviewed its different approaches and contrasted their characteristics on multiple dimensions such as knowledge representation learning, consideration of forgetting behavior, and architecture design. Moreover, we presented a chronological flow for deep learning approaches that showed how later methods extended former ones for a cohesive understanding of limitations and research directions in these approaches. 

Moreover, we introduced a detailed review of relevant KT datasets that have been used by the literature while describing their characteristics, limitations, and contribution points. Additionally, we provided a curated summary for all the reported performance results on the covered datasets for key KT methods.

Finally, we discussed various application areas for knowledge tracing to show its potential in addressing 
human and machine teaching domains. Furthermore, we highlighted future research directions that could push the boundaries of the current knowledge tracing methods by harnessing data from multiple modalities, learning with weak or no supervision, or following an interactive reinforcement learning paradigm to overcome cold start challenges.

 \bibliographystyle{ACM-Reference-Format}
  \bibliography{sample-base}

\end{document}